\documentclass[11pt]{article}

\usepackage[T1]{fontenc}
\usepackage[utf8]{inputenc}
\usepackage{lmodern}
\usepackage[margin=1in]{geometry}
\usepackage{microtype}
\usepackage{amsmath,amssymb,amsthm,mathtools,bm}
\usepackage{booktabs,array,tabularx}
\usepackage{enumitem}
\usepackage[numbers,sort&compress]{natbib}
\usepackage[hidelinks]{hyperref}
\usepackage[nameinlink,noabbrev]{cleveref}
\usepackage{xcolor}
\usepackage{caption}
\usepackage{xurl}
\usepackage[section]{placeins}

\usepackage{graphicx}

\setlist[enumerate]{leftmargin=.5in}
\setlist[itemize]{leftmargin=.5in}

\allowdisplaybreaks
\numberwithin{equation}{section}

\theoremstyle{plain}

\theoremstyle{definition}

\theoremstyle{remark}

\crefname{theorem}{Theorem}{Theorems}
\Crefname{theorem}{Theorem}{Theorems}
\crefname{proposition}{Proposition}{Propositions}
\Crefname{proposition}{Proposition}{Propositions}
\crefname{lemma}{Lemma}{Lemmas}
\Crefname{lemma}{Lemma}{Lemmas}
\crefname{corollary}{Corollary}{Corollaries}
\Crefname{corollary}{Corollary}{Corollaries}
\crefname{remark}{Remark}{Remarks}
\Crefname{remark}{Remark}{Remarks}
\crefname{assumption}{Assumption}{Assumptions}
\Crefname{assumption}{Assumption}{Assumptions}
\crefname{equation}{equation}{equations}
\Crefname{equation}{Equation}{Equations}
\crefname{section}{Section}{Sections}
\Crefname{section}{Section}{Sections}
\crefname{subsection}{Section}{Sections}
\Crefname{subsection}{Section}{Sections}
\crefname{table}{Table}{Tables}
\Crefname{table}{Table}{Tables}

\newcommand{\chla}{\mathrm{CHL\text{-}a}}

\title{How Much Hyperspectral Information Does Chlorophyll Retrieval Really Need?}
\author{
  \normalsize Abed Hammoud$^{1,*}$, Xuerong Sun$^{2}$, Bianca Champenois$^{3}$, Robert J.W. Brewin$^{2}$\\[0.8em]
  \begin{minipage}{0.96\textwidth}
    \centering\small
    $^{1}$ Civil and Environmental Engineering, Princeton University, Princeton, New Jersey 08540, USA\\[0.4em]
    $^{2}$ Centre for Geography and Environmental Science, Department of Earth and Environmental Sciences, Faculty of Environment, Science and Economy, University of Exeter, Exeter, Cornwall, UK\\[0.4em]
    $^{3}$ High Meadows Environmental Institute, Princeton University, Princeton, New Jersey 08540, USA
  \end{minipage}\\[0.6em]
  \small $^*$Corresponding author: \texttt{ah1389@princeton.edu}
}
\date{}

\hypersetup{
  pdftitle={How Much Hyperspectral Information Does Chlorophyll Retrieval Really Need?},
  pdfauthor={Abed Hammoud, Xuerong Sun, Bianca Champenois, Robert J.W. Brewin}
}

\begin{document}

\maketitle

\begin{abstract}
Satellite ocean color algorithms translate water-leaving radiance into ecological information at spatial and temporal scales that cannot be achieved by field sampling alone. 
One important variable derived from water-leaving radiance is chlorophyll-a concentration ($\chla$), a widely used indicator of phytoplankton biomass and physiology. 
Empirical retrieval algorithms are commonly used for $\chla$ estimation, but their performance can vary across sensors and optically diverse waters.
The Plankton, Aerosol, Cloud, ocean Ecosystem (PACE) mission provides unprecedented spectral resolution, expanding the visible spectral information available for ocean color retrievals and raising a practical algorithm design question: 
how can this fine resolution spectrum be used to derive the next generation of interpretable $\chla$ retrieval algorithms? 
We use symbolic regression to identify sparse equations that estimate $\log_{10}(\mathrm{CHL\text{-}a})$ from multi- and hyperspectral remote sensing reflectances. 
The analysis first tests the standard multispectral ocean color algorithm (OC3--OC6) inputs to ask whether symbolic regression recovers standard band ratio structure, then extends the search to hyperspectral PACE-like reflectances. 
The best expression discovered achieved a held out root mean square deviation of 0.253 in log$_{10}$ $\chla$, compared with 0.307 for a fitted OC6 polynomial on the same split.
Ablations showed that the predictive skill of hyperspectral models is almost 12\% better than the best standard ocean color retrieval models in root mean squared deviation against held out data. 
When evaluated by environmental regime, differences were larger for high-chlorophyll samples, which often represent turbid water conditions. 
Results support symbolic regression as a tool for proposing interpretable ocean color equations, and the new family of ocean color models pinpoints the most significant reflectance wavelengths.
\end{abstract}


\section{Introduction}

Ocean color remote sensing is a fundamental component of the Global Ocean Observing System (GOOS) that helps monitor upper-ocean biological variability over regional to global scales \cite{McClain2009Decade, Groom2019SatelliteOceanColour}. 
Satellite sensors measure the spectral radiance signal leaving the ocean after interaction with water, phytoplankton, mineral particles, colored dissolved organic matter, and the atmosphere. 
After atmospheric correction, remote sensing reflectance, $R_{rs}(\lambda)$, provides the basic optical input to many derived products, including chlorophyll-a ($\chla$), diffuse attenuation coefficients, and inherent optical properties \citep{nasa_obpg_rrs}. 
$\chla$ is a widely used satellite product for measuring phytoplankton biomass because it is a dominant phytoplankton pigment \cite{Sathyendranath2023}. 
It is also used to monitor phytoplankton physiology \cite{Behrenfeld2016}. 
Long-term global ocean color records of $\chla$ support studies of productivity, blooms, and biogeochemical variability \citep{nasa_obpg_product_suites, nasa_obdaac}.

For empirical ocean color retrievals, $\chla$ is estimated by fitting a relationship between in situ $\chla$ measurements and spectral inputs derived from $R_{rs}$, typically reflectance ratios at selected wavelengths.
The retrieval problem is challenging because the observed spectrum is influenced by several water constituents, measurement conditions, and processing uncertainties, so $\chla$ is difficult to estimate directly from $R_{rs}$.
As a result, an algorithm may perform well overall but still produce large errors in specific optical environments \citep{morel1977_ocean_color, werdell2018_iop_review}. 
Although evaluation cannot eliminate the ambiguity caused by multiple water constituents, carefully curated in situ observations spanning diverse optical regimes can reveal where this ambiguity leads to retrieval failure. 
Rigorous quality control and satellite matchup procedures further reduce measurement, processing, and spatiotemporal collocation errors, helping distinguish deficiencies in the retrieval algorithm from uncertainties in the in situ data \citep{werdell2005_biooptical, bailey2006_multisensor, valente2016_essd, valente2019_essd}.

Standard empirical $\chla$ algorithms address this problem by assuming that, within the waters represented by the training dataset, optically active constituents covary with $\chla$ such that the concentration can be represented by a low-dimensional empirical relationship. 
The ocean color (OCx) algorithm family uses a maximum blue-to-green band ratio, where the numerator is the largest $R_{rs}$ value among selected blue—bands, while its denominator is $R_{rs}$ at a green band. 
Because chlorophyll pigments absorb relatively strongly at blue wavelengths, blue reflectance generally decreases relative to green reflectance as $\chla$ increases. 
Selecting the maximum among several blue bands preserves this spectral contrast as the most responsive band changes across concentration regimes.
Related ocean color product families use low-dimensional band-ratio or band-difference for the target product \citep{oreilly1998_seawifs, oreilly2019_ocx, nasa_obpg_product_suites}. 
These algorithms remain attractive because their variables and coefficients can be inspected. 
Their limitations are equally clear: a fixed ratio form may not capture all information present in hyperspectral observations \cite{Wolanin2016SpectralBands, Vandermeulen2017SpectralSampling, Torrecilla2011ClusterAnalysis, Cael2023IndependentQuantities, SATHYENDRANATH2017125}, and empirical coefficients can be sensitive to the training database, sensor band set, and water-type composition \cite{BlondeauPatissier2014OceanColorReview}.
Moreover, polynomials often fail to generalize near the boundaries of the input range on which they were trained.

Modern machine learning retrievals offer an alternative by fitting more flexible nonlinear mappings from spectra and ancillary information to geophysical variables \cite{Tiwari2023}. 
Neural network (NN) products and comparisons are now part of operational and research ocean color workflows, including the Ocean and Land Colour Instrument (OLCI) CHL NN comparison product provided with Sentinel-3 products \citep{copernicus_olci_products, eumetsat_ocean_colour}. 
Similarly, probabilistic models, derived following similar modeling paradigms as that of ocean color algorithms, have also highlighted the strength of Bayesian methods in capturing uncertainties in matchup datasets \cite{Craig2019, Hammoud2025, Hammoud2026}.
Such models can be useful, but their flexibility often comes at the cost of direct interpretability. 
For scientific algorithm development, a strong prediction is not enough because users need to know which spectral features matter, when a model extrapolates, and whether a compact equation can be implemented and evaluated across sensors.

The Plankton, Aerosol, Cloud, ocean Ecosystem (PACE) mission changes the scale of the design problem. 
The Ocean Color Instrument (OCI) on PACE provides dense hyperspectral observations, 100s of reflectances, in the visible domain relative to conventional multispectral sensors, approximately 10 reflectances \citep{werdell2019_pace, nasa_pace_oci, meister2024_oci}. 
High-resolution dense spectra provide the opportunity to exploit absorption, scattering, and fluorescence-related structure, but they also make manual equation design more tedious. 
An exhaustive search over possible wavelength combinations, ratios, differences, nonlinear transforms, and auxiliary variables is not practical if the goal is still a concise and interpretable equation.

Symbolic regression offers a principled way to search large parametric spaces and derive analytical expressions describing the relationship between independent and dependent variables in functional form \cite{Tang2012SymbolicRegression, Chami2002GeneticProgramming, Huang2021CoastalZoneImager}. 
Previous studies, particularly \citet{Tang2012SymbolicRegression}, have already demonstrated that symbolic regression can produce explicit $\chla$ retrieval equations from multispectral band-ratio inputs.
The novelty of the present study lies in extending and evaluating the approach across conventional OC3--OC6 multispectral configurations and PACE-like hyperspectral observations. 
We further quantify the tradeoff between predictive accuracy and expression complexity using held-out data, compare the discovered equations against OC$x$ models refitted on identical data partitions and regression baselines, examine performance across environmental and $\chla$ regimes, and demonstrate their application to satellite scenes while diagnosing reflectances outside the training domain. 
These methodological and validation advances, detailed in the four contributions below, distinguish the present framework from earlier proof-of-concept applications of symbolic regression to ocean color retrieval.
Rather than prescribing a single functional form, symbolic regression uses an evolutionary, also known as genetic, programming search over computational trees, balancing prediction error against expression complexity. 
Generally, symbolic regression returns a Pareto set of candidate equations with different accuracy--complexity tradeoffs \citep{cranmer2023_pysr}. 
Concise analytic models are derived by these genetic programming approaches, but also that results require careful validation, robustness checks, and domain interpretation \citep{lacava2021_srbench, makke2024_sr_review}. 
In this manuscript, we therefore treat symbolic regression outputs as empirical candidate algorithms that, once validated, might be linked to physical laws.

This work makes four main contributions:
\begin{enumerate}
\item It uses symbolic regression to test whether interpretable ocean color algorithms can be learned directly from reflectance data, beginning with the multispectral OC3--OC6 setting to assess whether standard band ratio structures emerge from the data.
\item It extends this framework to PACE-like hyperspectral reflectances and shows that dense spectral information can be compressed into sparse, closed-form chlorophyll algorithms using only a small number of wavelengths.
\item It performs controlled ablation experiments across legacy multispectral bands, coarser spectral resolutions, selected spectral windows, and full PACE-like sampling to quantify how much predictive skill is gained from increased spectral density.
\item It evaluates robustness across chlorophyll ranges, water types, SST regimes, and optically complex conditions, while also showing that adding SST provides limited improvement relative to reflectance-only models.
\end{enumerate}
Together, these contributions demonstrate that symbolic regression can discover compact and interpretable ocean color algorithms while clarifying where hyperspectral information improves upon standard ocean color approaches and where models remain vulnerable.

The remainder of the manuscript is organized as follows. Section~\ref{sec:data} describes the data products and spectral feature construction. 
Section~\ref{sec:methods} describes the target transformation, OCx baselines, PySR search, validation design, diagnostics, and PACE scene workflow. 
Section~\ref{sec:results} presents experiments and results of the study. 
Section~\ref{sec:discussion} interprets the scientific implications and limitations, and summarizes the main conclusions.

\section{Data}
\label{sec:data}

\subsection{Data products}
\label{ssec:data}

Two complementary bio-optical datasets were used in this study. 
The first contains multispectral matchups between in situ \(\mathrm{CHL}\text{-}a\) measurements and satellite remote sensing reflectances, \(R_{rs}\), and is based on the global compilation of \citet{Valente2022}. 
The second consists of concurrent in situ hyperspectral \(R_{rs}\) and \(\mathrm{CHL}\text{-}a\) observations, and is an expanded dataset of the one used in \citet{Sun2025RSE}, and is referred to throughout this study as the hyperspectral dataset.

The multispectral experiments used the extensive global bio-optical database of \citet{Valente2022}, which integrates observations from 27 independently assembled datasets.
The complete Valente et al. compilation contains measurements of \(\mathrm{CHL}\text{-}a\) obtained by fluorometry or high-performance liquid chromatography, remote-sensing reflectance, and several inherent and apparent optical properties, such as phytoplankton absorption, \(a_{ph}\), particle backscattering, \(b_{bp}\), and diffuse attenuation coefficient \(K_d\) to name a few. 
The observations were collected from international repositories and long-term oceanographic programs, including SeaBASS \citep{Fargion2002}, NOMAD \citep{WERDELL2005}, MERMAID \citep{barker2008mermaid}, ICES, ARCSSPP, BIOCHEM, BODC, COASTColour \citep{Nechad2015}, MAREDAT \citep{peloquin2013}, and SeaDataNet, as well as MOBY \cite{clark2003}, BOUSSOLE, AERONET-OC, HOT, GeP\&CO, AMT, AWI, BARENTSSEA, BATS, CALCOFI, CCELTER, CIMT, ESTOC, IMOS, PALMER, TPSS, and TARA. 
A detailed description of the individual data sources and their processing is provided by \citet{Valente2022}.

For the present analysis, we used the matchups between in situ \(\mathrm{CHL}\text{-}a\) and multispectral satellite \(R_{rs}\). 
The satellite observations were acquired by the Sea-viewing Wide Field-of-view Sensor (SeaWiFS), the Moderate Resolution Imaging Spectroradiometer (MODIS), the Medium Resolution Imaging Spectrometer (MERIS), the Visible Infrared Imaging Radiometer Suite (VIIRS), and the Ocean and Land Colour Instrument (OLCI). 
The matchups have broad global coverage across both open-ocean and coastal waters. 
This dataset was used for the OC3--OC6 experiments because it represents the conventional multispectral satellite setting in which empirical ocean-color algorithms are developed and evaluated.

The second dataset is a global compilation of concurrent in situ hyperspectral  \(R_{rs}\) and \(\mathrm{CHL}\text{-}a\) measurements. 
We refer to this compilation as the hyperspectral dataset, compiled from previous studies \citet{Sun2025RSE, Sun2025Frontiers}. 
The version used here is broader than the open-ocean subset analyzed by \citet{Sun2025RSE, Sun2025Frontiers}, because samples from turbid and other optically complex waters were retained to represent a wider range of conditions. 
The observations were obtained from NASA SeaBASS, PANGAEA, the British Oceanographic Data Centre, and individual published datasets \cite{garaba2011pangaea759690, zielinski2013pangaea811777, xi2021pangaea930087, bracherrottgers2022pangaea946394, bracher2019pangaea898920, bracher2014pangaea848589, brachercheah2022pangaea946368, bracheretal2018pangaea884526, peekenmurawski2016pangaea867473, bracher2015pangaea848585, bracher2017pangaea879226, bracher2017pangaea879225, tayloretal2011pangaea819099, bracherwiegmann2019pangaea899043, bracherwiegmann2022pangaea944796, krameretal2021pangaea937536, valenteetal2022, brewinetal2023, lewisarrigo2020}. 

Remote sensing reflectances are measured using above-water hyperspectral radiometers following established ocean optics protocols \citep{Mueller2003}. 
Measurements having a native spectral resolution of 2~nm or finer were interpolated onto a regular 1~nm wavelength grid spanning 400--700~nm. 
Thus, the tabulated 1~nm reflectances represent spectrally interpolated measurements rather than observations acquired at a native 1~nm resolution. 
Spectra containing spuriously high reflectance values, defined as \(R_{rs}>0.15\), or spectral quality assurance scores below 0.80 were removed \citep{Wei2016ReflectanceQuality}. 
Raman scattering corrections were applied following \citet{Lee2013UVVisiblePenetration}.

Concurrent \(\mathrm{CHL}\text{-}a\) concentrations were measured using high-performance liquid chromatography or in vitro fluorometry. 
When more than one measurement method was available for the same observation, the HPLC measurement was retained preferentially. 
Daily sea surface temperature (SST) was obtained from the NOAA Optimum Interpolation Sea Surface Temperature version~2 product and matched to each field observation using the nearest spatial grid cell on the corresponding sampling date.

The final hyperspectral dataset contains observations collected between 1991 and 2019, together with sampling date, geographic coordinates, measurement depth, SST, data source and cruise, total \(\mathrm{CHL}\text{-}a\), and wavelength-resolved \(R_{rs}\). 
\(\mathrm{CHL}\text{-}a\) concentrations range from 0.012 to 130.165~\(\mathrm{mg\,m^{-3}}\), with a median of 0.800~\(\mathrm{mg\,m^{-3}}\). 
The continuous visible spectra make this dataset suitable for constructing PACE-like reflectance inputs and for performing controlled spectral-resolution experiments. 
It was therefore used for the hyperspectral symbolic regression analysis, the 2.5, 5, and 10~nm spectral-resolution ablations, the corresponding multispectral band subsets, and the analyses of model performance across \(\mathrm{CHL}\text{-}a\) and SST regimes.

\subsection{PACE-like and multispectral input construction}
\label{ssec:input_construction}

For the Valente et al.\ matchup dataset, the matched satellite \(R_{rs}\) values at 412, 443, 490, 510, 560, and 665 nm were used directly to construct the OC3, OC4, OC5, and OC6 multispectral input systems \citep{oreilly2019_ocx}. 
Symbolic regression was then applied separately to each band set (reflectances and not MBR) to determine whether the search recovered the compact blue-to-green ratio structure and the polynomial family of the established OCx family or identified alternative expressions with improved accuracy at comparable complexity.

For the PACE-like hyperspectral experiments, the interpolated 1~nm in situ spectra were transformed into several observing-system representations. The primary PACE-like representation used the target band centers
\begin{equation}
\lambda_j = 402.5 + 2.5j~\mathrm{nm},
\qquad j=0,\ldots,118,
\end{equation}
yielding 119 reflectance predictors spanning 402.5--697.5~nm.
At each target wavelength, the predictor was calculated by averaging the available 1~nm \(R_{rs}\) values using a normalized Gaussian weighting: 
\begin{equation}
\overline{R}_{rs}(\lambda_c) =
\frac{\sum_i R_{rs}(\lambda_i)\exp\left[-\frac{(\lambda_i-\lambda_c)^2}{2\sigma^2}\right]}
{\sum_i\exp\left[-\frac{(\lambda_i-\lambda_c)^2}{2\sigma^2}\right]}.
\label{eq:gaussian_inputs}
\end{equation}
where, $\lambda_i$ spans the wavelengths within $\lambda_c \pm 2.5\,\mathrm{nm}$, and $\sigma = \mathrm{5}/(2\sqrt{2\ln 2}) \approx 2.12\,\mathrm{nm}$.
This representation approximates the 2.5~nm spectral sampling and nominal 5~nm bandwidth of the PACE OCI over the visible wavelength range covered by the in situ dataset. 
The same spectra of the hyperspectral dataset were used to construct the 5~nm and 10~nm, and multispectral OC3--OC6 representations employed in controlled spectral ablation experiments. 
The OC3--OC6 subsets constructed from the hyperspectral dataset are distinct from the primary Valente et al.\ multispectral experiment.

Observations with missing or non-finite values in any predictor required by a given representation were excluded from the corresponding experiment. 
Construction of the complete 119-band PACE-like representation retained 1,299 observations from the original 6,265 samples. 
Data partitioning was performed separately for each experiment. 
The PACE-like analysis used training, validation, and unseen-test subsets, where model parameters were fitted using the training data, candidate expressions were selected using validation root mean squared difference, and the test subset was evaluated only after model selection. 
The observing-system ablation used the common 85/15 split reported in \Cref{fig:ablation}. 
The robustness experiment used ten outer 90/10 splits, with model selection performed using a validation partition.

\section{Methods}
\label{sec:methods}

\subsection{Target variable and evaluation}

All primary models were trained to predict
\begin{equation}
y_i = \log_{10}(C_i),
\end{equation}
where $C_i$ is total surface $\chla$ in mg m$^{-3}$. 
Metrics are therefore reported in log$_{10}$ $\chla$ space. 
The log transformation follows common practice in chlorophyll algorithm evaluation \cite{Campbell1995LognormalBioOptical}, where concentration ranges span orders of magnitude and residuals are more interpretable in multiplicative terms.
Note that we also employ the symbolic regression algorithm to fit a linear model of $C_i$, however, the results underperformed those of the $\log_{10}$ transformed ones and consequently were omitted from the manuscript. 

Model skill was assessed by comparing predicted and observed chlorophyll-a concentrations in logarithmic space, consistent with standard practice in ocean color algorithm evaluation \citep{Brewin2015, Hammoud2025}. 
Let $y_i = \log_{10}(C_i)$ and $\hat{y}_i = \log_{10}(\hat{C}_i),$ denote the observed and predicted chlorophyll-a values for validation sample (i), respectively. 

The primary error metric was the mean squared difference, $\mathrm{MSD} = \frac{1}{N}\sum_{i=1}^{N}(\hat{y}_i-y_i)^2,$ together with the root mean squared difference, $\Psi = \sqrt{\frac{1}{N}\sum_{i=1}^{N}(\hat{y}_i-y_i)^2},$ where \(\Psi\) denotes the root mean squared difference in \(\log_{10}\) chlorophyll space. 
The mean prediction bias was computed from the residuals \(e_i=\hat{y}_i-y_i\). 
We report the absolute bias, $\delta = \left|\frac{1}{N}\sum_{i=1}^{N}e_i\right|,$ and the unbiased root mean squared difference, $\Delta = \sqrt{\Psi^2-\delta^2}$.
Together, \(\delta\) and \(\Delta\) separate systematic offset from the residual scatter about the mean error.

Linear association between predictions and observations was quantified using the Pearson correlation coefficient with \(r^2\) reported as an additional measure of explained variance. 
To characterize departures from the one-to-one relationship, we also fit the linear relation $y_i = S\hat{y}_i + I + \varepsilon_i,$ where (S) and (I) are the slope and intercept, respectively. 
A model with perfect agreement would therefore have (S=1), (I=0), zero $\delta$, and zero $\Psi$. 
These diagnostics were used both to compare candidate symbolic regression equations and to summarize the performance of the selected equation on the validation set.

\subsection{Ocean Color baseline models}

The maximum band ratio ocean color baselines were included to provide a direct comparison with established ocean color algorithm structure \cite{oreilly2019_ocx}. 
For each OCx variant, we computed a maximum-band ratio
\begin{equation}
\rho_x =
\log_{10}
\left(
\frac{\max_{\lambda \in B_x} R_{rs,\lambda}}
{D_x}
\right),
\end{equation}
where $B_x$ is the blue band set for the chosen OCx definition and $D_x$ is either a green reflectance or the OC6 mean green/red denominator. 

The OCx family is an x$^{th}$ order polynomial function of $\rho$.
For example, a $4^{th}$ order polynomial was then fit on the training rows:
\begin{equation}
\hat{y} = a_4\rho^4 + a_3\rho^3 + a_2\rho^2 + a_1\rho + a_0.
\end{equation}
For OC3--OC5, $D_x$ is the 560 nm reflectance. 
OC6 uses the same maximum-blue numerator as OC5 but replaces the denominator with the mean of 560 and 665 nm reflectance. 
The polynomial coefficients were fit on each training split. 
Published fixed OCx coefficients were not used in the primary comparison, because the goal was to compare the discovered model with the best regressed standard functional family on the same dataset.

\begin{table}[htbp]
\small
\centering
\caption{Multispectral and hyperspectral OCx definitions used in this study \citep{oreilly2019_ocx}.}
\label{tab:ocx_definitions}
\begin{tabularx}{\linewidth}{@{}llXX@{}}
\toprule
Algorithm & Numerator bands & Multispectral denominator & Hyperspectral denominator \\
\midrule
OC3 & max(443, 490) & 560 & 555 \\
OC4 & max(443, 490, 510) & 560 & 555 \\
OC5 & max(412, 443, 490, 510) & 560 & 555 \\
OC6 & max(412, 443, 490, 510) & mean(560, 665) & mean(555, 670) \\
\bottomrule
\end{tabularx}
\end{table}

\subsection{Symbolic regression}

Symbolic regression was used to identify compact analytical mappings from remote-sensing reflectance spectra to \(\mathrm{CHL}\text{-}a\). 
We used PySR \citep{cranmer2023_pysr}, a genetic-programming symbolic regression package in which each candidate model is represented as an expression tree. 
The leaves of the tree are input variables or numerical constants, and the internal nodes are mathematical operators. 
During the search, candidate expressions are generated through mutation, recombination, simplification, and coefficient optimization. 
This differs from linear or polynomial regression because both the algebraic structure of the model and its numerical coefficients are learned from the data.

For each observing system experiment, the input vector consisted of the available remote-sensing reflectance values, \(R_{rs}(\lambda)\), and the response variable was the logarithm of \(\mathrm{CHL}\text{-}a\).
The symbolic regression search space included the binary operators ${+, -, \times, \div},$ and the unary operators ${x^2, x^3, x^{-1}, x^{-2}, \log_{10}(x), \exp(x), |x|}$.
Thus, arbitrary exponentiation and square root transformations were not included in the search space. 
PySR was trained using the elementwise squared error loss, $(\hat{y}-y)^2,$ so that the empirical loss corresponds to the MSD in \(\log_{10}\) space. 

The symbolic regression runs used 5000 iterations, 50 populations, a maximum expression size (the total number of nodes in an expression tree) of 60, a maximum tree depth of 60, a batch size of 512, and a maximum of \(5\times 10^6\) expression evaluations. 
The parsimony coefficient (a multiplicative penalty added to the loss function during the evolutionary search to discourage complex mathematical expressions) was set to 1, constants were assigned a complexity weight of 2, and randomization of fitted weights was enabled with weight-randomization factor 0.01.

Model complexity was quantified using PySR's expression-tree complexity, defined as the sum of node-specific complexity weights over all nodes in the expression. 
This complexity measure penalizes longer expressions and encourages sparse, interpretable equations that use only a small subset of the available wavelengths. 
PySR returns a set of candidate equations spanning a range of complexity--accuracy tradeoffs. 
Rather than selecting the final expression solely from the internal PySR score, we used a hold-out validation procedure. 
The data were randomly split into 85\% training and 15\% validation sets using a fixed random seed. 
PySR was fitted on the training set, after which the top 50 performing equations ranked by the symbolic regression score were re-evaluated on the validation set. 
Among equations with complexity not exceeding 50, the final model was chosen as the equation with the lowest validation mean squared difference, with validation \(r^2\) used as a secondary criterion. 
Validation performance was then summarized using $\Psi$, MSD, absolute bias, unbiased RMSD, correlation coefficient, \(r^2\), and the slope and intercept of the regression between predicted and observed \(\log_{10}(\chla)\).

\section{Results}
\label{sec:results}

\subsection{Symbolic regression as a bridge to OCx structure}

We first consider the conventional multispectral setting to determine whether symbolic regression recovers structures related to the established OCx family when provided with the same spectral information. 
In particular, the results presented in this section are based on the multispectral dataset of \citet{Valente2022}.
Separate searches were performed using the reflectance bands associated with OC3, OC4, OC5, and OC6 as input. 
Because OC4 is among the most widely used members of this family, its discovered complexity trajectory is examined in detail before extending the comparison across all four observing systems. 
This experiment provides a controlled test of whether a broader algebraic search reproduces the band-ratio logic underlying OCx or identifies alternative compact mappings from reflectance to $\chla$.

\begin{figure}
    \centering
    \includegraphics[width=0.95\linewidth]{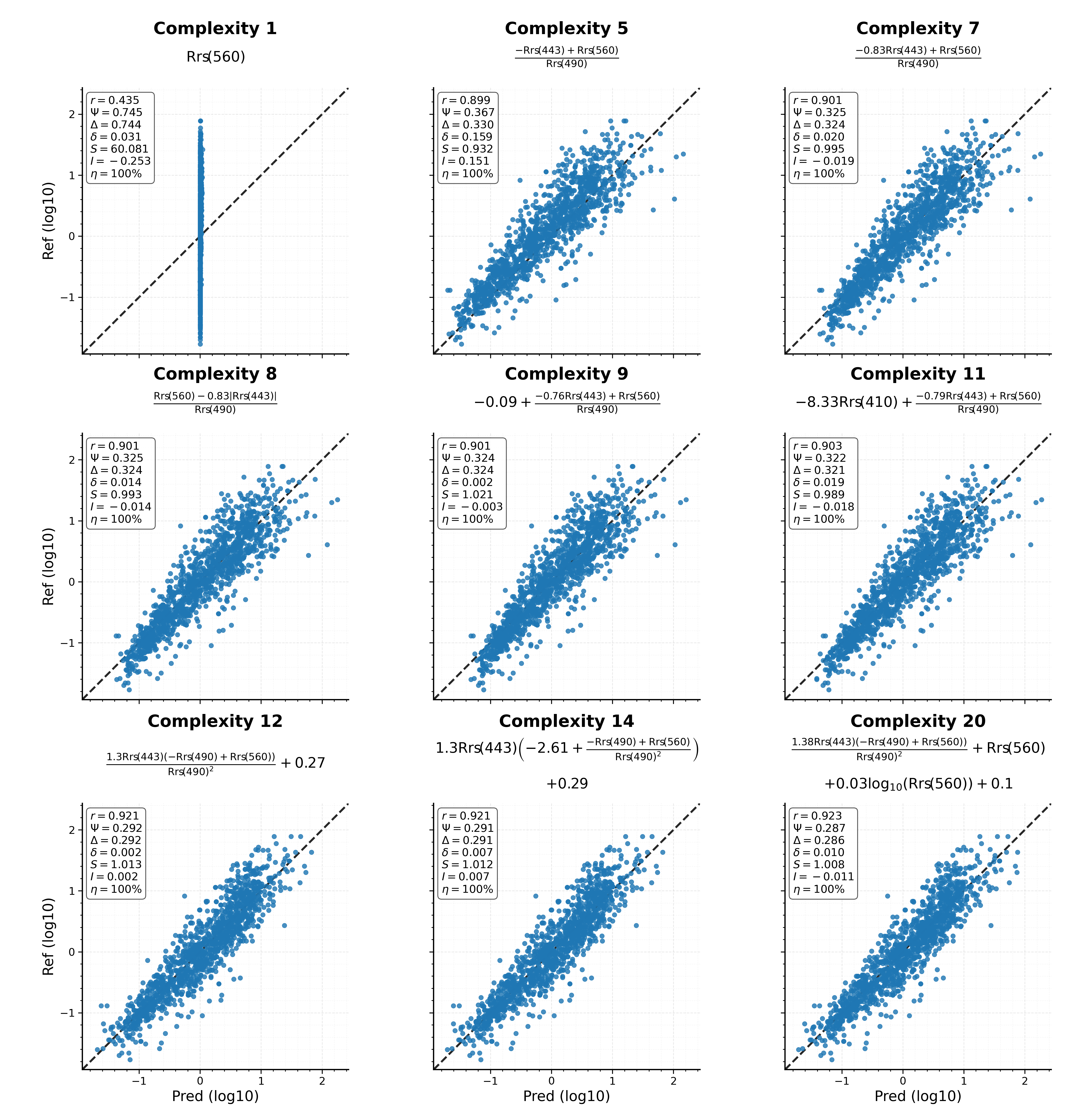}
    \caption{Evolution of representative OC4 symbolic regression candidates with increasing expression complexity. 
    Each panel shows predicted versus reference ($\log_{10}$) $\chla$ for a retained candidate, with the corresponding display equation and evaluation metrics. 
    The dashed line denotes perfect one-to-one agreement.
    Increasing complexity first introduces a normalized blue--green reflectance contrast and subsequently adds fitted coefficients, offsets, and additional spectral terms. 
    Most of the reduction in RMSD occurs by complexities 7--12, whereas further increases in complexity provide comparatively small improvements. 
    Equations are rounded for display, while metrics were calculated using full-precision predictions.}
    \label{fig:oc4_complexity}
\end{figure}

\Cref{fig:oc4_complexity} shows representative OC4 symbolic regression candidates spanning expression complexities from 1 to 20. 
The complexity 1 model, which depends only on $R_{rs}(560)$, has limited predictive skill, with r=0.435 and an RMSD of 0.745. 
Introducing a normalized contrast among the 443, 490, and 560~nm bands at complexity 5 produces the largest initial improvement, increasing the correlation to $r$=0.899 and reducing the RMSD to 0.367. 
This expression nevertheless retains an absolute bias of 0.159. Small changes to the coefficient multiplying $R_{rs}(443)$, followed by the introduction of an additive offset, substantially reduce this bias (at complexity 7 or 8). 
At complexity 9, the discovered model achieves an RMSD of 0.324, an absolute bias of 0.002, and a regression slope of 1.021.

The error decreases more gradually beyond this point. 
Adding $R_{rs}(410)$ at complexity 11 reduces the RMSD only slightly, from 0.324 to 0.322. 
A second appreciable improvement occurs at complexity 12, where the expression introduces a different normalization involving $R_{rs}(490)^{2}$ and reaches an RMSD of 0.292 with $r$=0.921. 
Increasing the complexity from 12 to 20 lowers the RMSD by only a further 0.005, from 0.292 to 0.287 at the cost of significant complexity. 
All retained expressions provide finite predictions for the complete evaluation set. 
Results show that most of the predictive gain is obtained by relatively compact ratio-like expressions, whereas subsequent algebraic additions provide progressively smaller improvements.

The structure of the discovered candidates is also noteworthy. 
Rather than producing unrelated expressions as complexity increases, the search repeatedly combines contrasts among blue and green reflectances with normalization by another visible band. 
The coefficients and offsets evolve to correct bias and improve calibration, but the principal spectral structure remains stable. 
Symbolic regression therefore recovers an empirical family closely related to the spectral-ratio reasoning of OCx, despite being allowed to search over a substantially broader collection of mathematical expressions.

\Cref{fig:valente_bridge} extends this comparison to the OC3--OC6 band systems. 
The left column presents fitted degree-matched OCx polynomials, while the right column presents the validation-selected symbolic expressions subject to complexity C$\leq$10. 
For OC3, the fitted polynomial and symbolic model obtain RMSD values of 0.280 and 0.287, respectively. 
The corresponding values for OC4 are 0.303 and 0.324. 
The symbolic expressions become marginally more accurate for the larger band systems: the OC5 symbolic model obtains an RMSD of 0.301, compared with 0.303 for the fitted polynomial, while the OC6 symbolic model obtains an RMSD of 0.288, compared with 0.295. 
Across the four symbolic models, correlations range from 0.901 to 0.916, slopes range from 0.963 to 1.021.

These results indicate that most of the predictive skill of the polynomial family can be retained using substantially fewer fitted numerical constants and an explicitly inspectable dependence on individual reflectances. 
The OC3--OC6 polynomials contain four to seven fitted coefficients, whereas the displayed symbolic models contain only one or two fitted numerical constants and use two or three reflectance bands. 
Such parsimony simplifies model implementation, facilitates inspection of the contribution of each band, and reduces the number of fitted parameters combating the risk of overfitting. 

The multispectral experiments therefore position symbolic regression as a bridge to established ocean-color algorithm structure: when spectral information is restricted to conventional bands, the search independently recovers compact contrast- and ratio-based expressions with skill comparable to that of fitted OCx polynomials. 
Having established this connection, we next relax the multispectral constraint and investigate whether PACE-like hyperspectral sampling supports similarly compact equations that exploit spectral information beyond the conventional OCx bands.

\begin{figure}[p]
    \centering
    \includegraphics[width=0.485\linewidth]{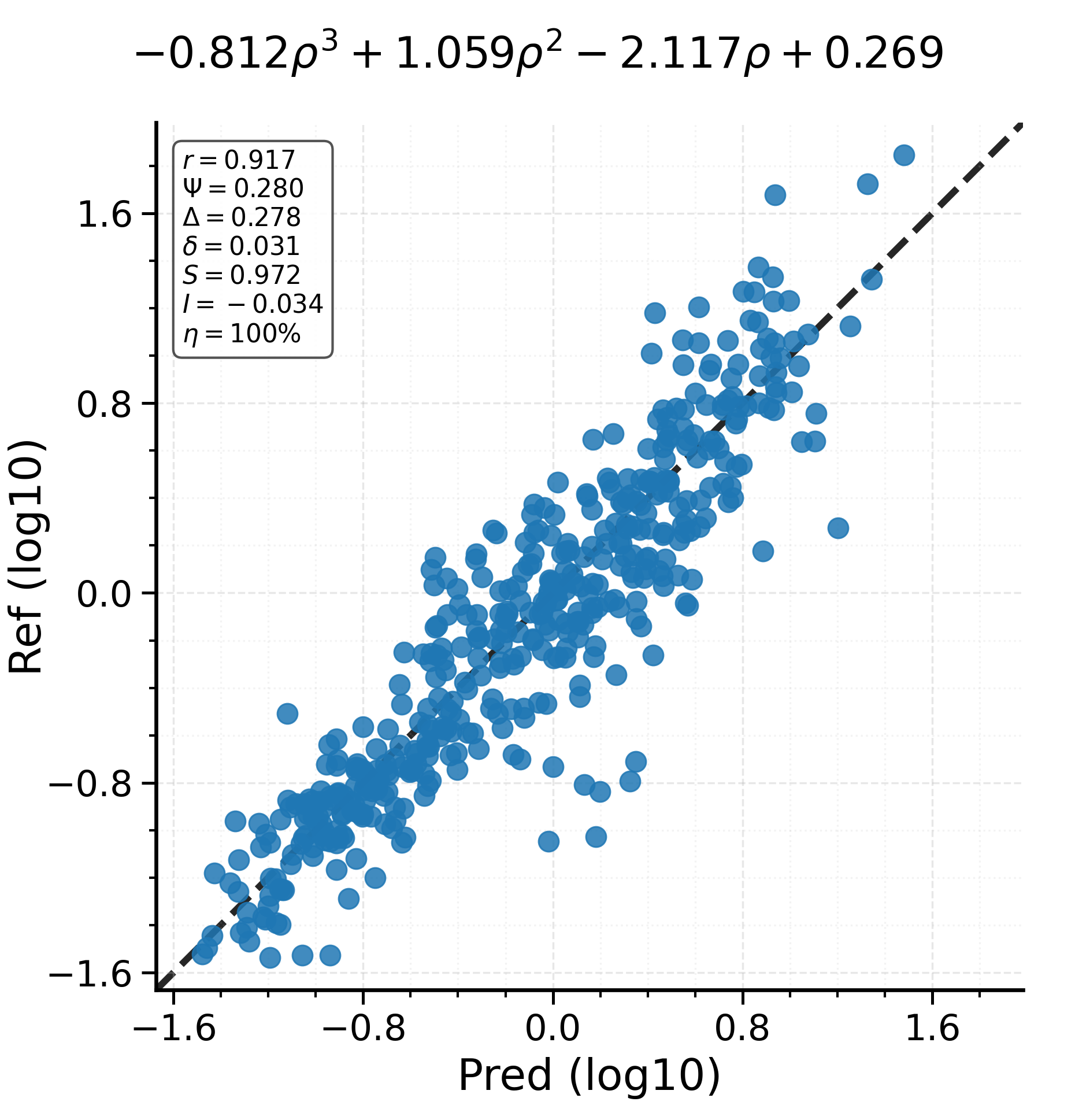}\hfill
    \includegraphics[width=0.485\linewidth]{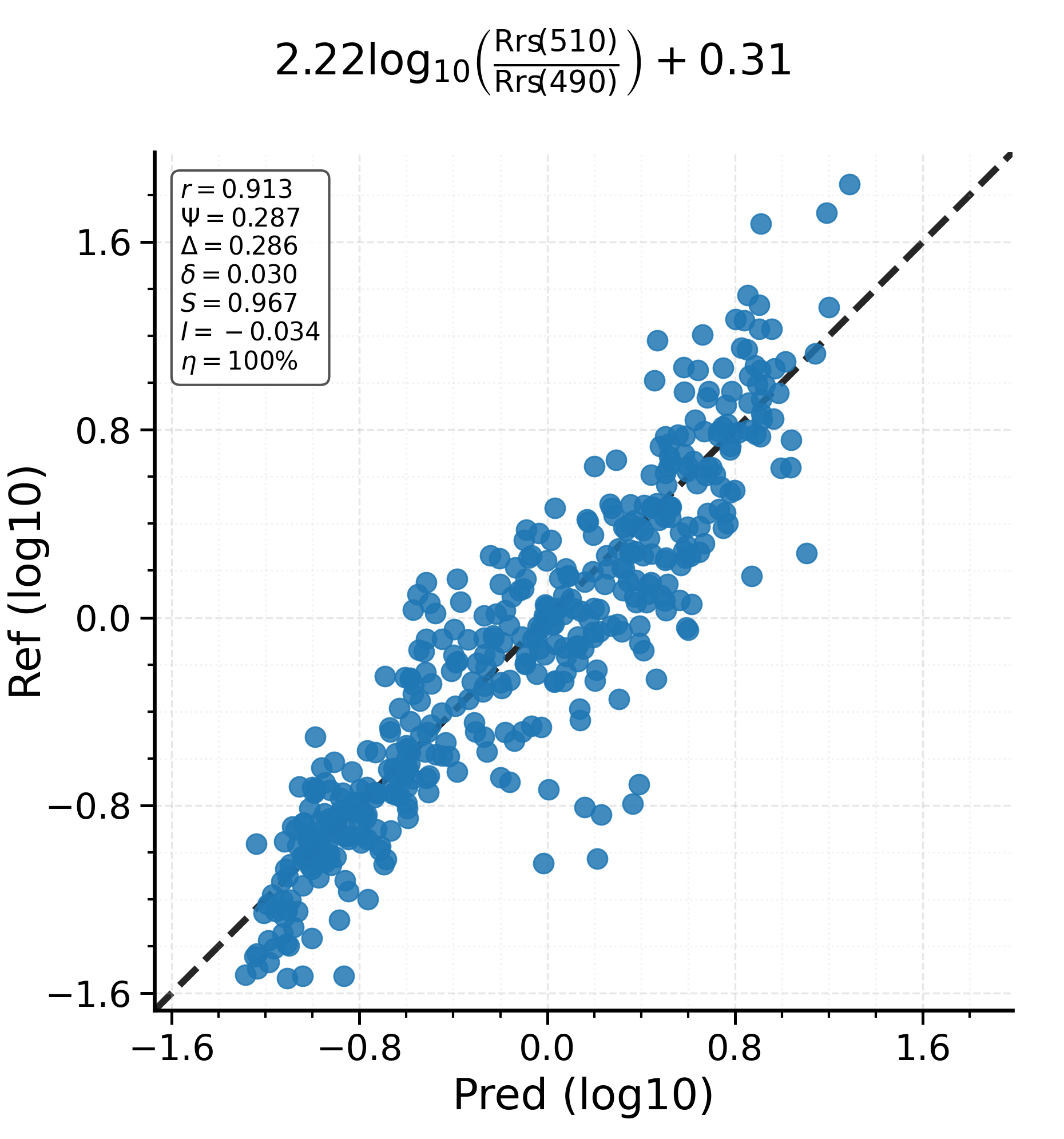}

    \vspace{0.4em}
    \includegraphics[width=0.485\linewidth]{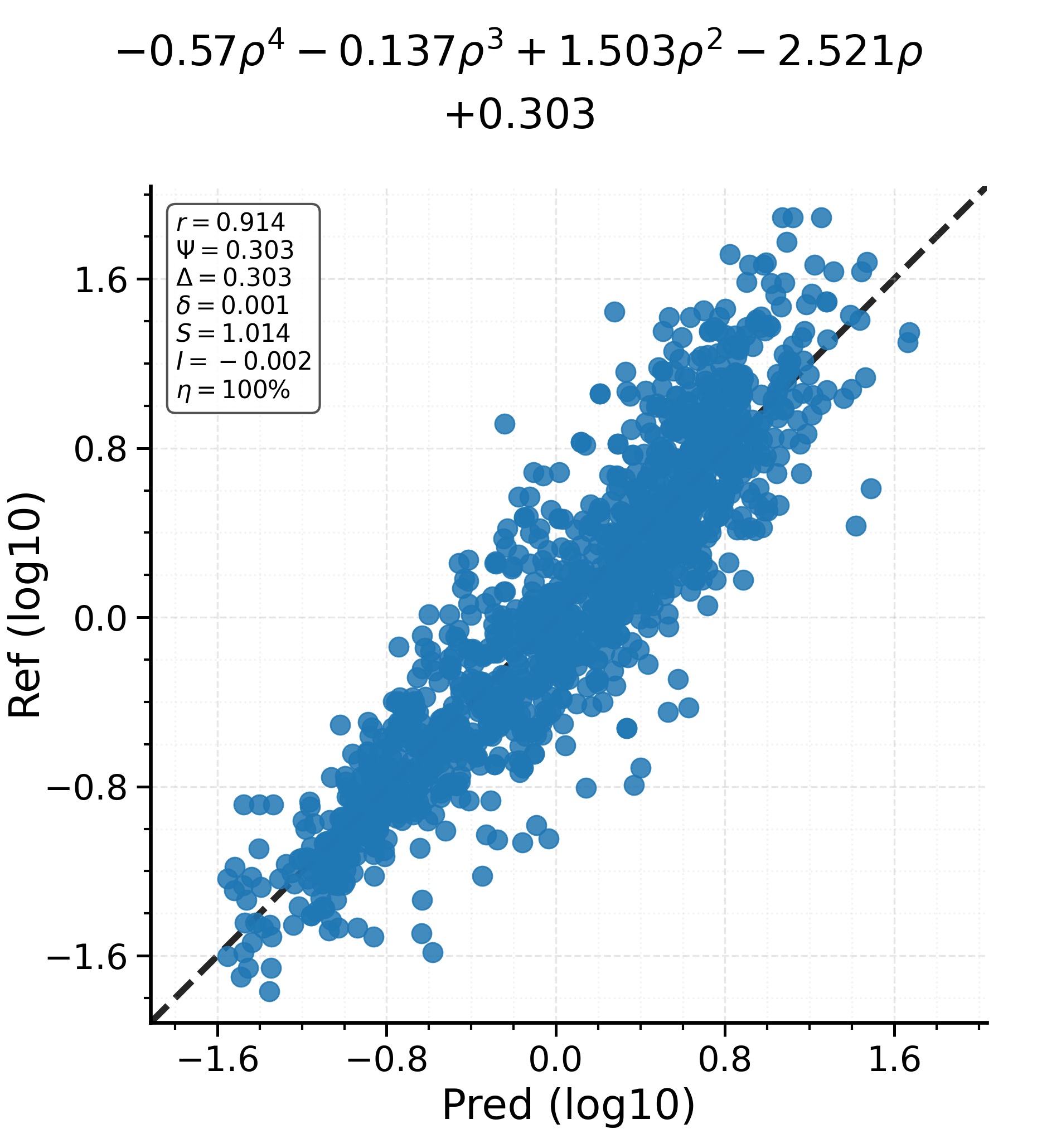}\hfill
    \includegraphics[width=0.485\linewidth]{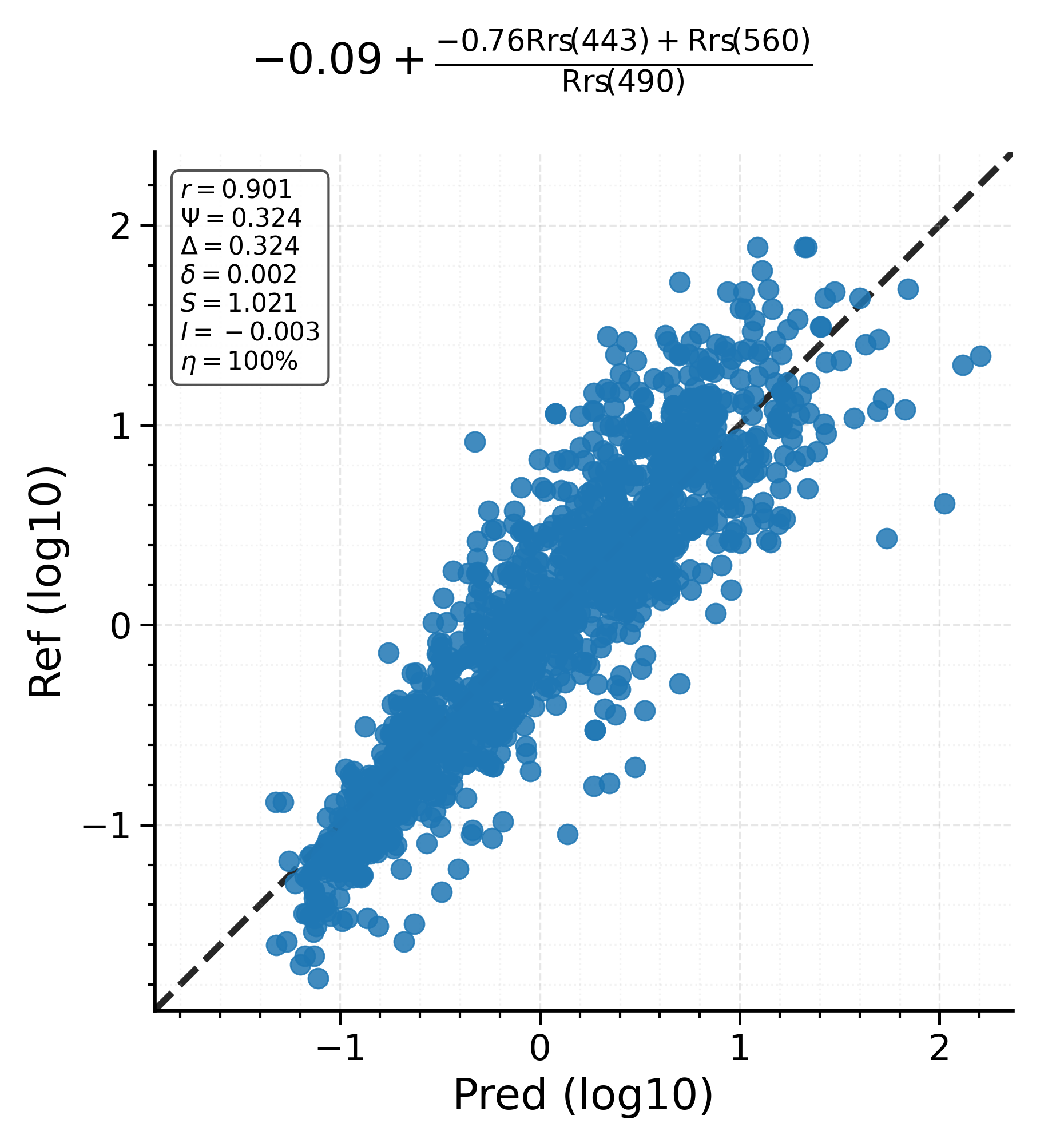}
    \caption{Degree-matched OCx LASSO polynomials (left) and validation-selected symbolic regression models with complexity $C\leq10$ (right) on the Valente dataset-based models: OC3 (top) and OC4 (bottom). Each pair uses the same reflectance columns and rows. Equations are rounded for display; metrics were computed from full-precision predictions. The metric blocks report correlation $r$, RMSD $\Psi$, unbiased RMSD $\Delta$, absolute bias $\delta$, reference-on-prediction slope $S$ and intercept $I$, and finite-prediction coverage $\eta$. Predictions are on the horizontal axis and references are on the vertical axis. The figure continues on the next page.}
    \label{fig:valente_bridge}
\end{figure}

\begin{figure}[p]
    \ContinuedFloat
    \centering
    \includegraphics[width=0.485\linewidth]{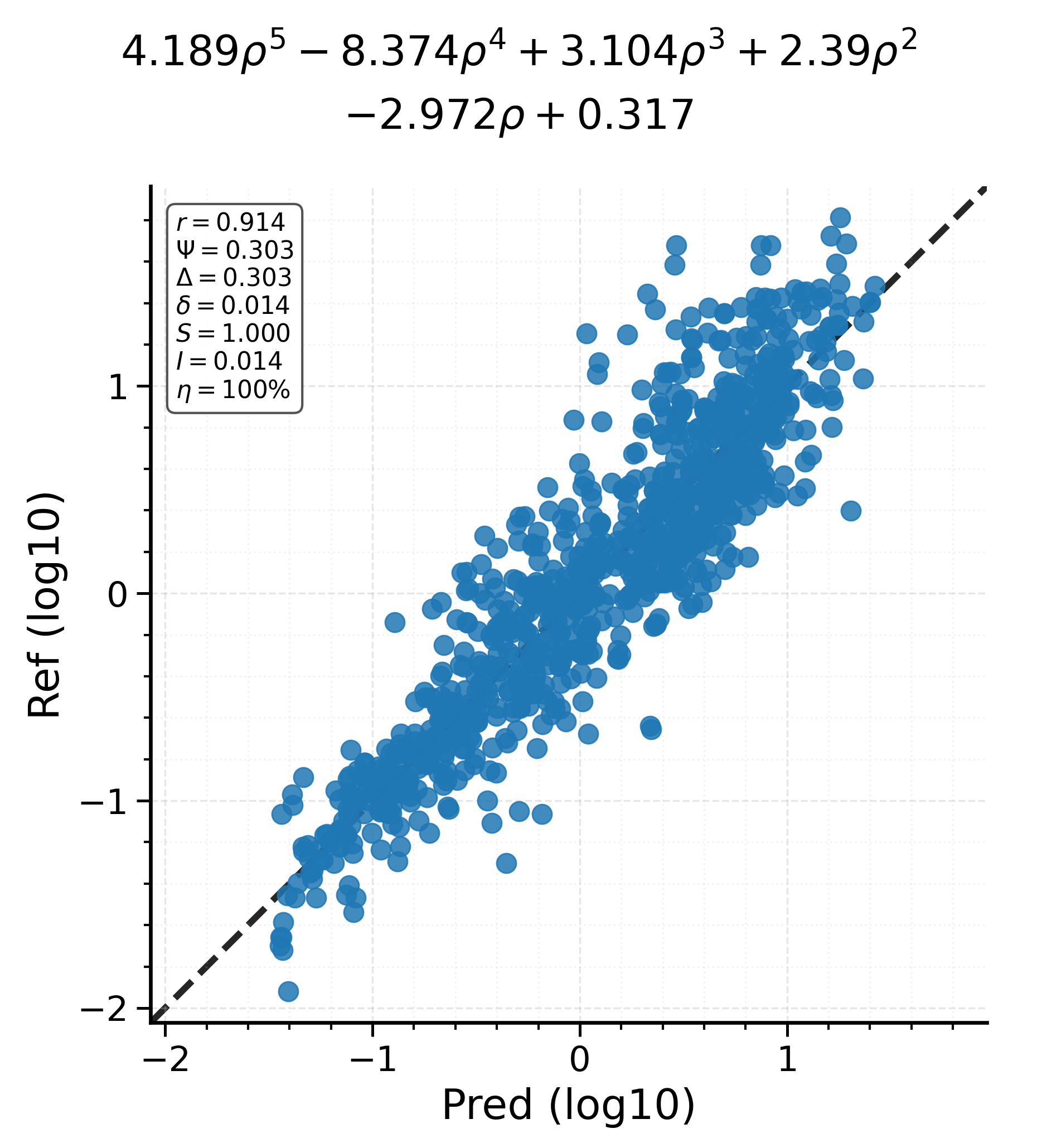}\hfill
    \includegraphics[width=0.485\linewidth]{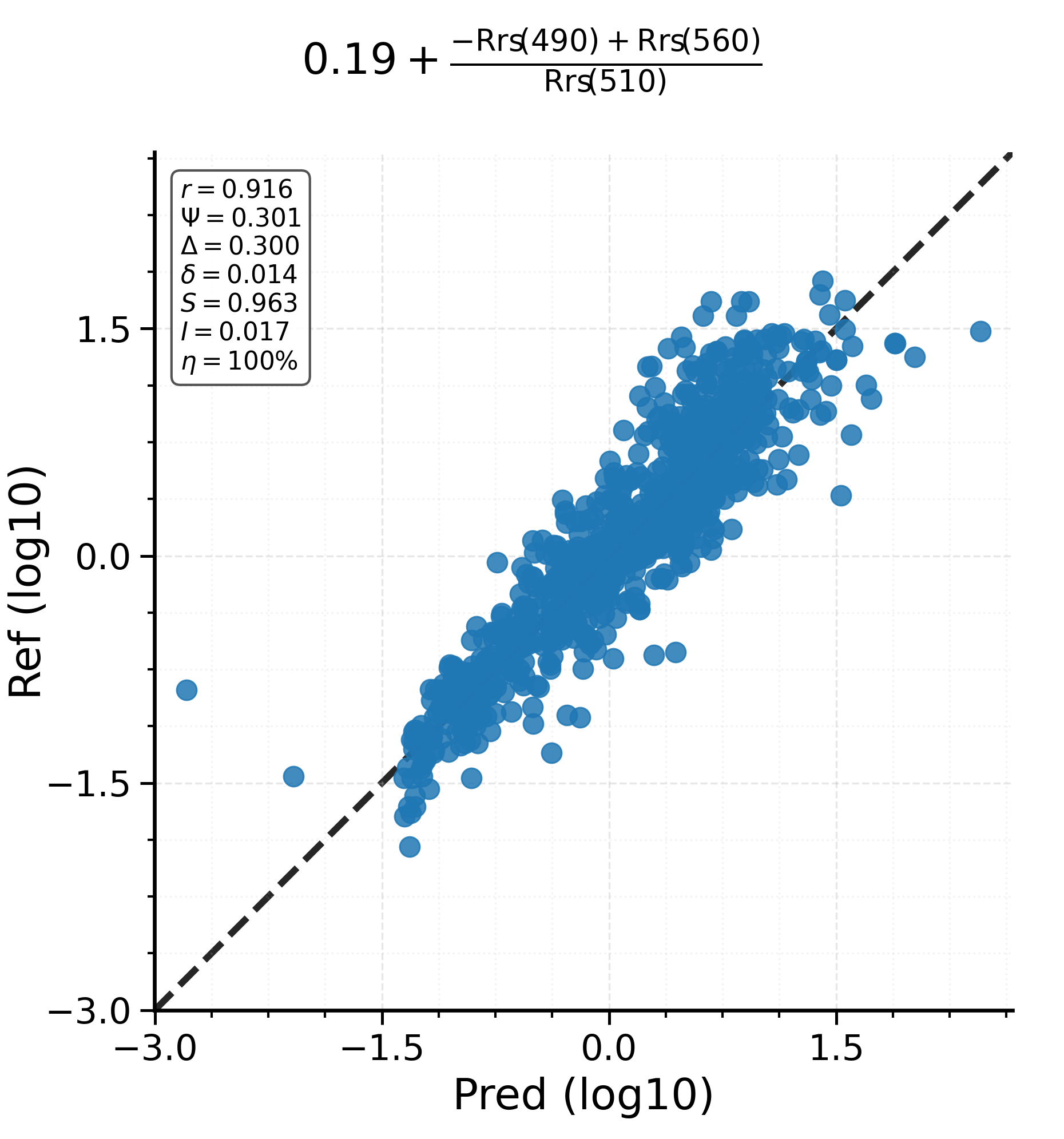}

    \vspace{0.4em}
    \includegraphics[width=0.485\linewidth]{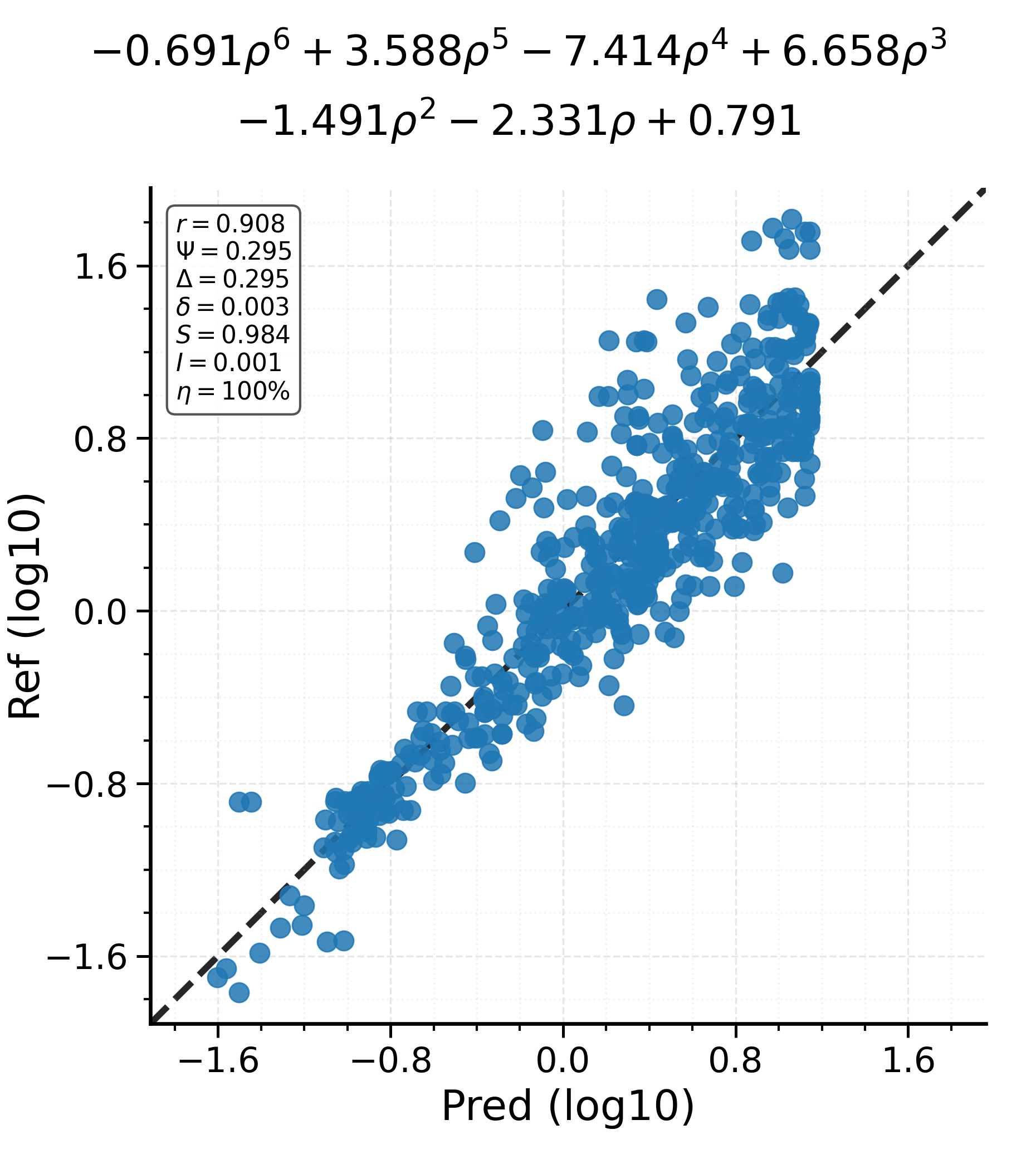}\hfill
    \includegraphics[width=0.485\linewidth]{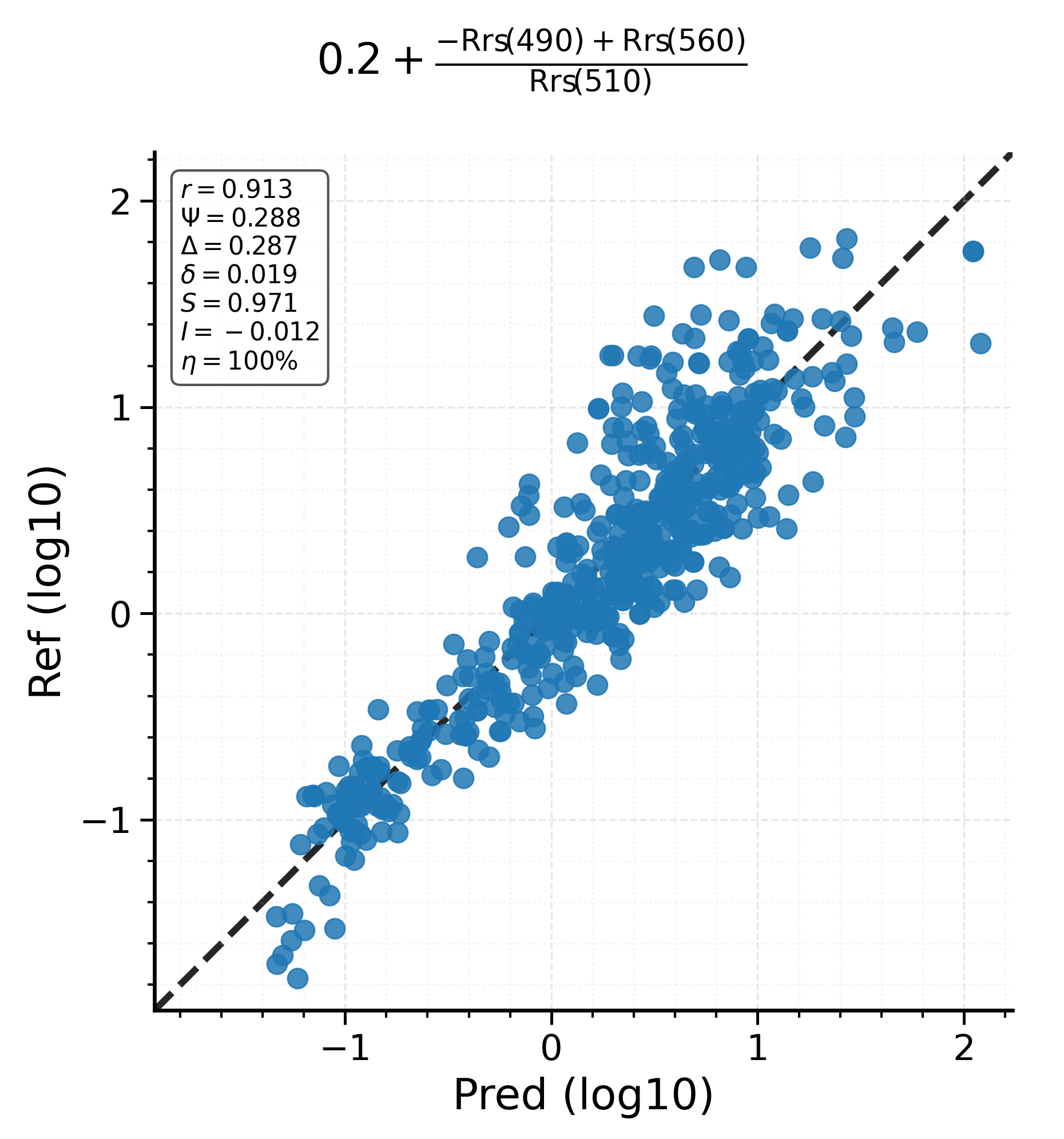}
    \caption[]{Valente dataset-based parity plots (continued): OC5 (top) and OC6 (bottom), with LASSO on the left and symbolic regression on the right. Metric definitions and axis orientation are as stated on the preceding page. Nominal wavelengths label pooled sensor roles and are not identical physical band centers for every sensor representation.}
\end{figure}

\subsection{PACE-like Hyperspectral Ocean Color Models}

The search is now expanded to the hyperspectral PACE-like reflectance set.
The OCI of PACE provides reflectances ranging from 315~nm in the ultraviolet to 895~nm in the near-infrared spectrum at 2.5~nm resolution (with bandwidths of 5 nm).
The in situ hyperspectral dataset provides reflectances ranging from 400~nm to 700~nm at 1~nm intervals, so we extract reflectances that resemble the PACE hyperspectral data, or at least a subset thereof.
Note that from this section onward, all the symbolic regression models were discovered using the hyperspectral dataset. 

The best equation discovered combines blue-green and red-edge reflectances. 
The equation uses only three of the 119 available constructed bands and retains a blue-green/red-edge ratio form: 
\begin{equation}
\widehat{y}_{\mathrm{PACE}}=2.51\log_{10}\!\left[\frac{\overline R_{\mathrm{rs}}(692.5)+\overline R_{\mathrm{rs}}(520)}{\overline R_{\mathrm{rs}}(490)}\right]-0.13,
\label{eq:PACE_selected}
\end{equation}
where $\overline R_{\mathrm{rs}}$ denotes the Gaussian-averaged input reflectance defined in \Cref{eq:gaussian_inputs}.

\begin{figure}[htbp]
    \centering
    \includegraphics[width=0.98\linewidth]{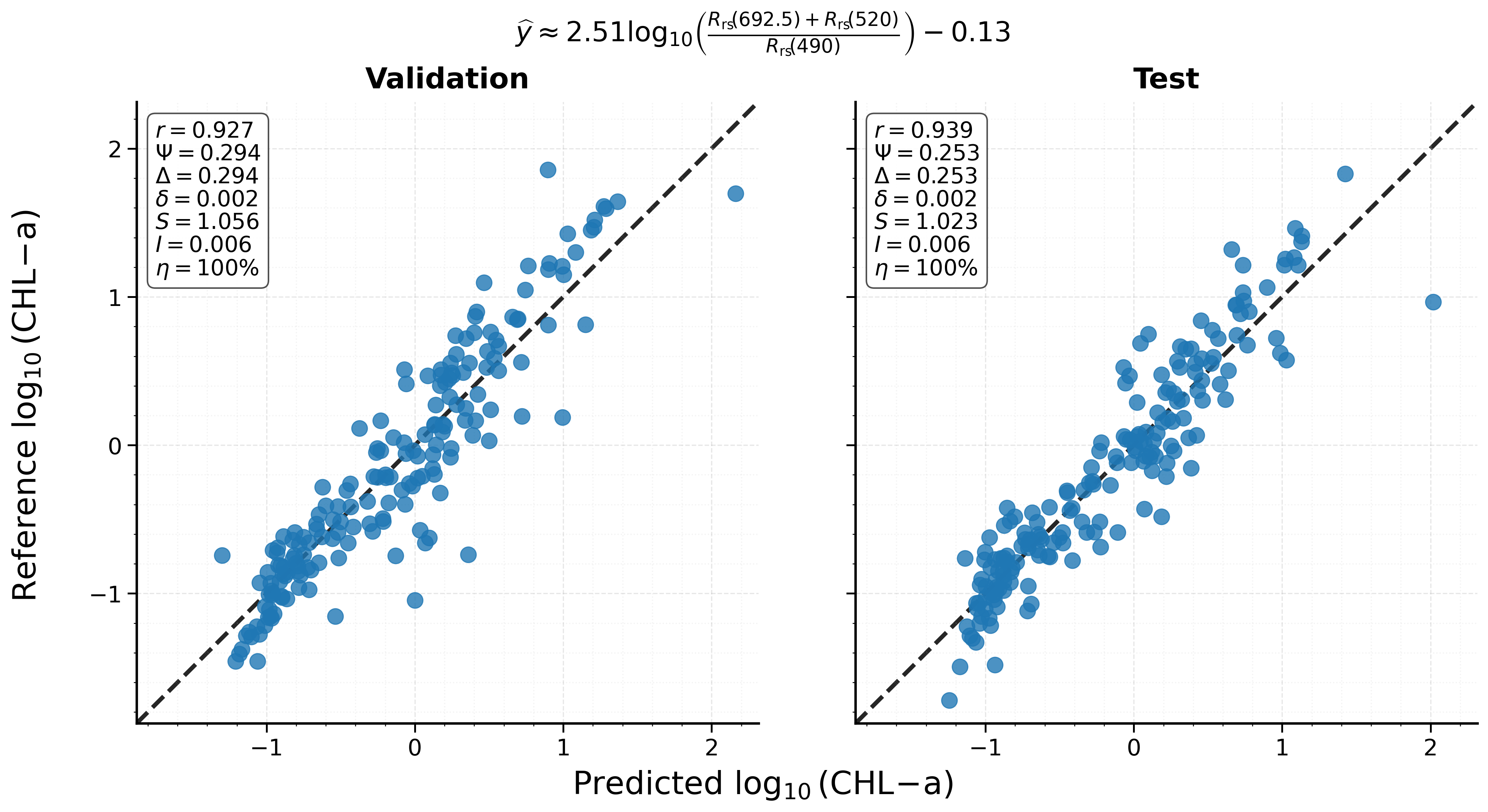}
    \caption{Validation-selection (left) and unseen test (right) parity plots for the complexity 8 expression. The equation was selected by minimum validation RMSD across three full-production searches after restricting retained candidates to complexity $C\leq10$. Test predictions were evaluated only after selection.}
    \label{fig:pace_parity}
\end{figure}

\Cref{fig:pace_parity} presents parity plots for the validation and test datasets, contrasting the predictions from the discovered model (\Cref{eq:PACE_selected}) with the in-situ counterpart.
The validation RMSD was 0.294, with $r=0.927$, based on 180 observations. 
The unseen test set yielded an RMSD of 0.253, an unbiased RMSD of 0.253, an absolute bias of 0.002, and $r=0.939$ (\cref{fig:pace_parity}). 
Results indicate that the discovered model generalizes well to the independent data, with reliable predictions and low RMSD values. 

Although the held out results demonstrate that a compact PACE-like expression can generalize beyond the data used for model selection, a single train-test partition cannot establish whether this behavior is robust to different data partitions or to the stochastic nature of the symbolic regression search. 
We therefore repeat the complete discovery procedure across multiple data partitions and independent symbolic regression searches.

\subsection{Statistical Significance of Discovered Models}

The present set of results relies on an operational setting in which a single data split is employed to first train the model, then test its generalizability on a reserved unseen subset. 
We now take a step further by investigating whether the reliability of the discovered family of models is statistically significant.
To do so, we perform 10 repetitions, each with a unique random seed initialization, of a 90/10 random data split to perform the symbolic regression on, and contrast the performance of the discovered models. 

\begin{figure}[htbp]
    \centering
    \includegraphics[width=0.92\linewidth]{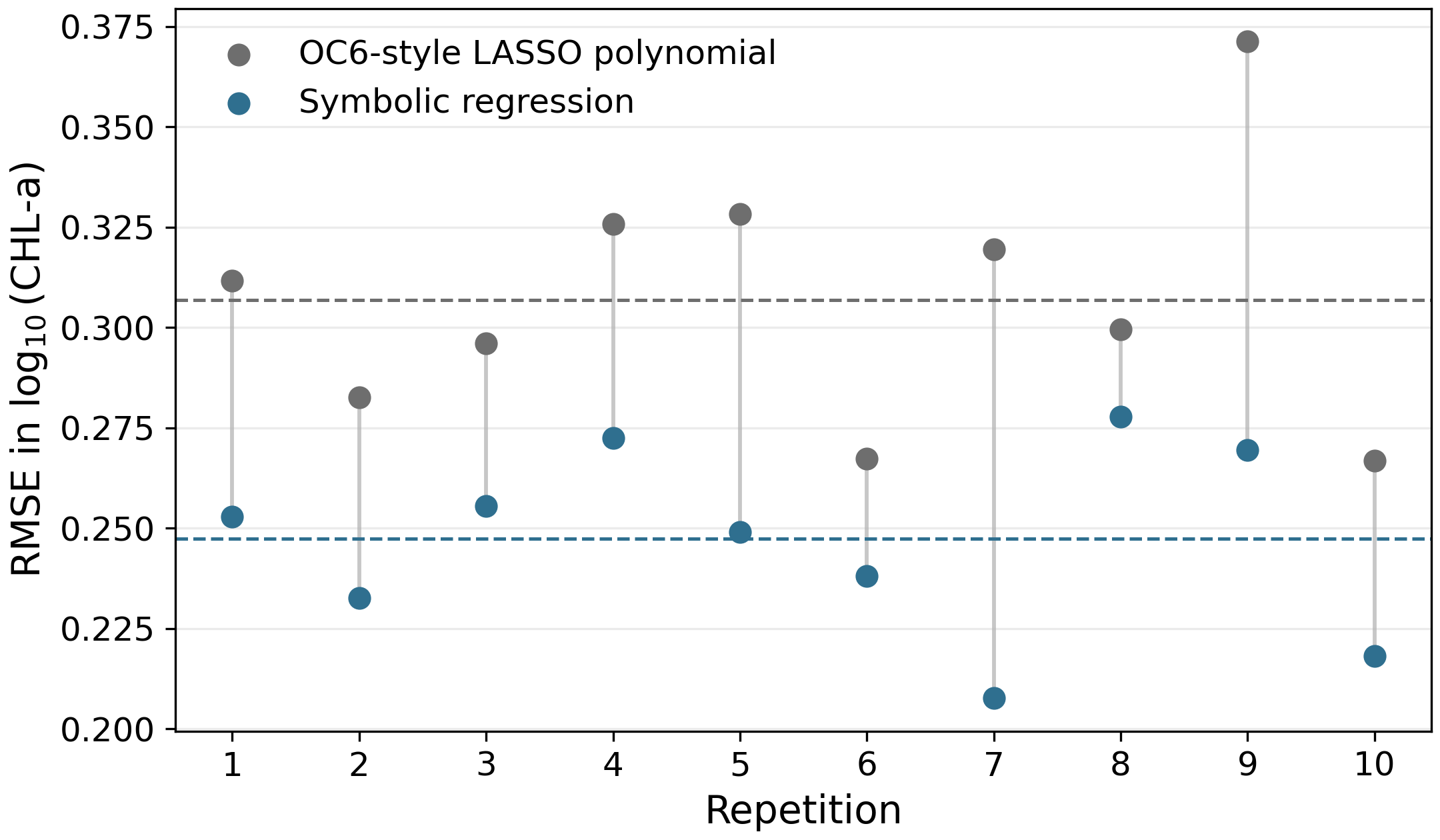}
    \caption{Test RMSD for ten 90/10 splits and a fixed random seed. Each blue point is the model selected from ten independent symbolic regression searches using the validation partition. The gray point is the OC6 LASSO polynomial refitted for the same split. Connecting lines identify paired results.}
    \label{fig:rep_rMSDs}
\end{figure}

The nested repeated experiment comprised 100 symbolic searches. 
The symbolic model was selected based on the validation RMSD conditioned on complexity less than or equal to 15, so that the discovered model is interpretable and within the same complexity as the OC6 model. 
\Cref{fig:rep_rMSDs} presents the RMSD of the OC6 and discovered model for 10 different random data splits and a fixed random seed.
The plot indicates that the symbolic regression can repeatedly produce ocean color models that result in lower RMSD values than the standard OCx ones. 

The average differences favored the model discovery algorithm, including their paired confidence intervals.
In particular, the discovered model results in a prediction RMSD of 0.247 with standard deviation 0.023, whereas the fitted OC6 polynomial resulted in a larger RMSD of 0.307 and a similar standard deviation of 0.032. 
The 95\% confidence intervals also indicate that the discovered models are generally more reliable than the fitted OC6 model, with respective ranges of \([0.231,\,0.264]\) and \([0.284,\,0.330]\). 

\begin{table}[!htbp]
\small
\centering
\caption{Repeated 90/10 full-feature PACE-like validation RMSD across 10 row-random split seeds. Values are in log$_{10}$ $\chla$ space.}
\label{tab:repeated_validation}
\begin{tabular}{lrrrr}
\toprule
Model & Mean & Std. dev. & Minimum--maximum & 95\% CI \\
\midrule
Gaussian PACE-like PySR & 0.247 & 0.023 & 0.208--0.278 & 0.231--0.264 \\
Gaussian OC6-LASSO & 0.307 & 0.032 & 0.267--0.371 & 0.284--0.330 \\
\bottomrule
\end{tabular}
\end{table}

\begin{figure}[htbp]
    \centering
    \includegraphics[width=0.95\linewidth]{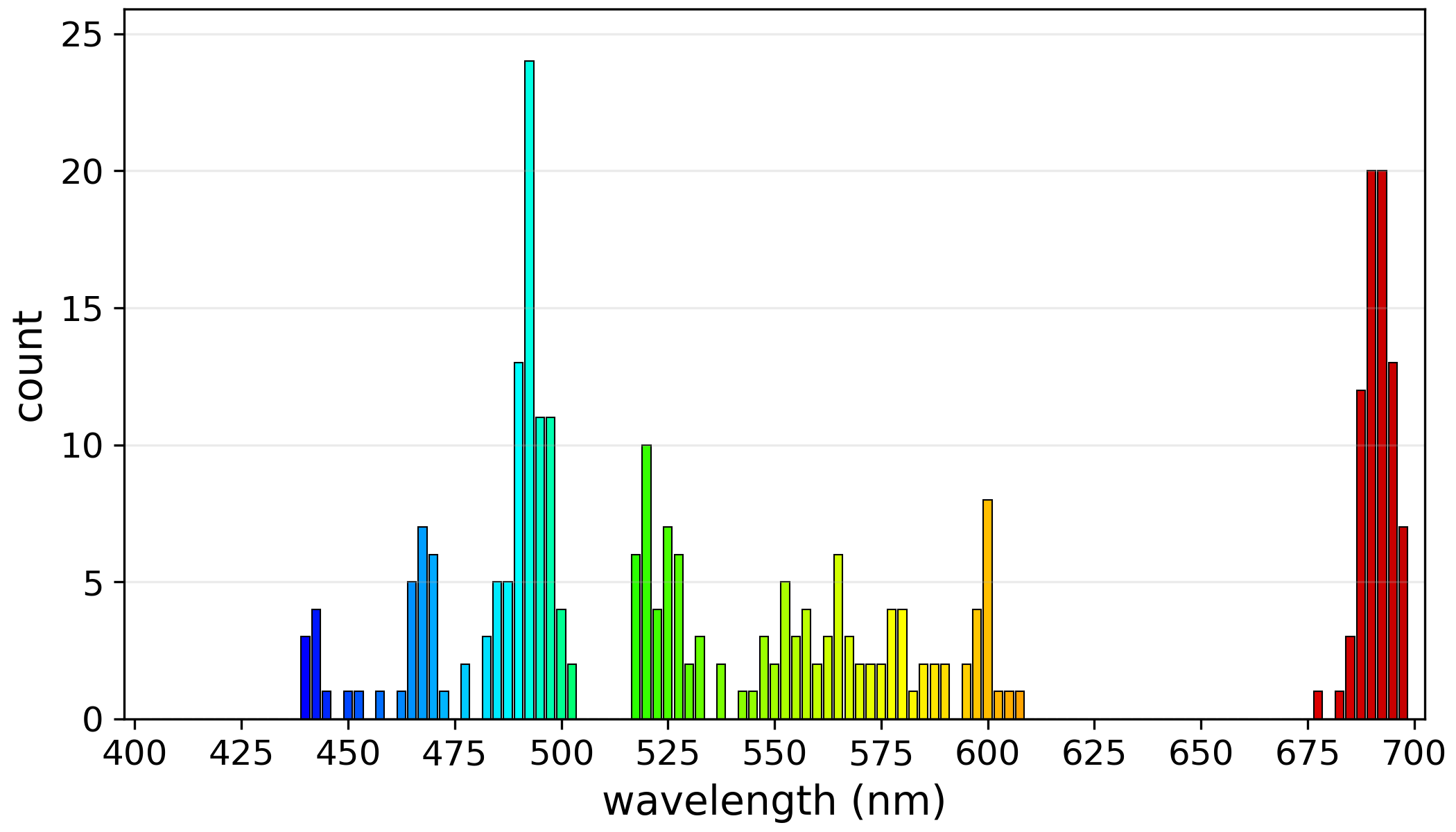}
    \caption{Count of wavelength occurrence among 100 selected equations in the separate archived 10-seed by 10-fold stability analysis. Bar colors denote the approximate visible-spectrum wavelength. A selected equation can contain more than one wavelength, so counts do not sum to 100. }
    \label{fig:rep_wavelengths}
\end{figure}

This result motivates the question: what bands were most significant to interpretable model discovery algorithm?
The frequency of appearance of each of the different reflectance wavelengths appearing in the discovered model was recorded for each of the best discovered models of the repetition of the experiment. 
\Cref{fig:rep_wavelengths} plots a barplot showing the count of the wavelengths employed by the sparse model discovery algorithm.
The plot indicates that 3 intervals of reflectances are most adopted by the model discovery algorithm. 
These wavelengths lie along the intervals in the blue-green, green and red/red-edge wavelengths, or in terms of wavelengths, these correspond to \([485,\ 510]\) nm, \([515,\ 525]\) nm and \([680, 700]\) nm, respectively. 
Furthermore, examining the structure of the discovered models revealed that the best models generally belong to the log-ratio family, suggesting the optimality of this family for ocean color modeling of $\chla$.

The repeated experiments establish that the performance advantage is not limited to one favorable split and that the selected expressions repeatedly draw from a small number of spectral regions. 
This raises a complementary observing system question: how much of the available hyperspectral information is actually required to obtain this improvement?

\subsection{Ablation across observation systems}

The observing system ablation asks whether denser spectral sampling would produce a performance gain again using all hyperspectral reflectances as input. 
To this end, we compare the performance of the OC3--OC6 models, fitted through regression, and model discovery using the same reflectances as OC3--OC6 as well as subsets of the hyperspectral dataset consisting of 2.5 nm, 5nm and 10 nm resolutions.

\begin{figure}[htbp]
    \centering
    \includegraphics[width=0.99\linewidth]{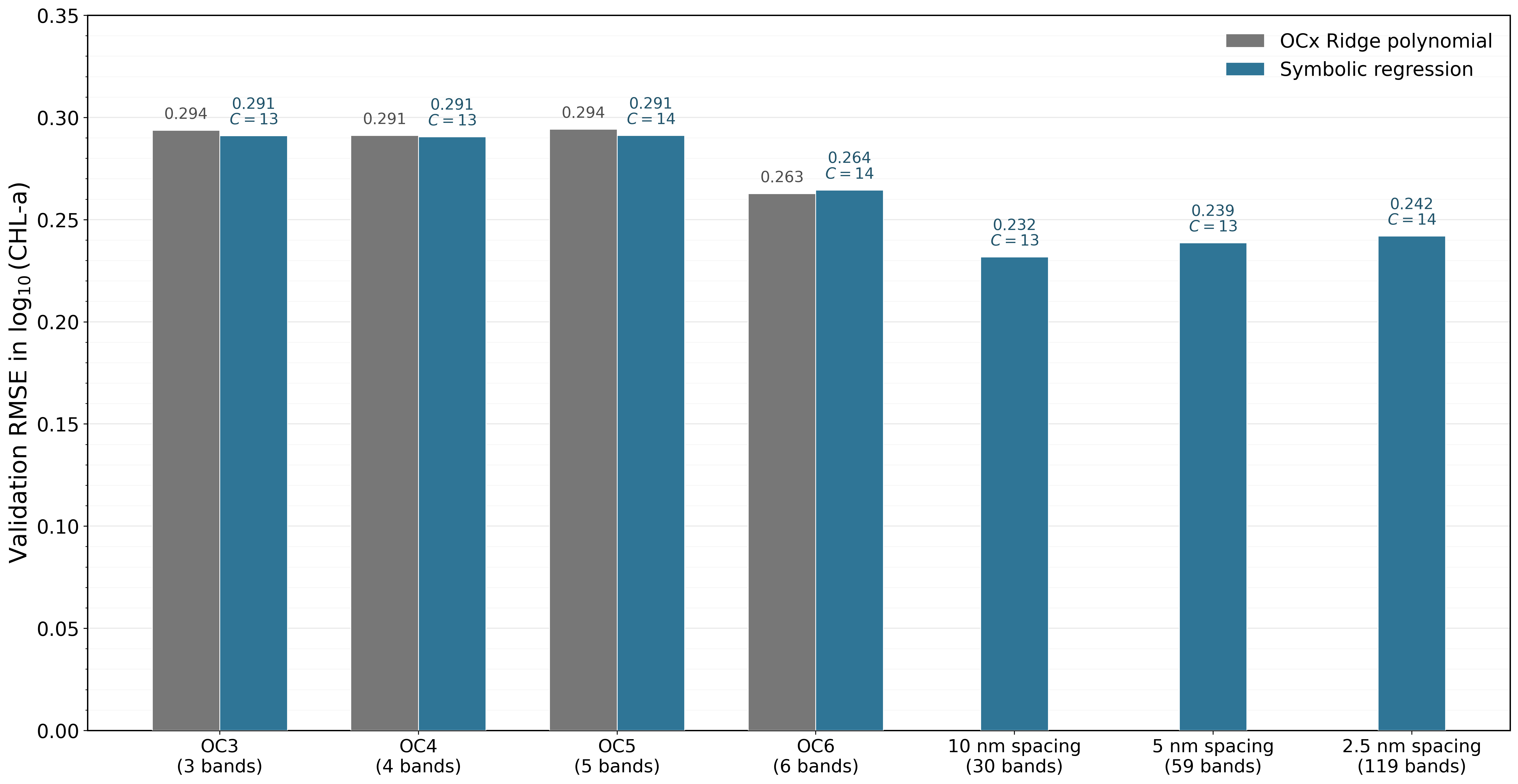}
    \caption{Validation-selection RMSD for OCx Ridge polynomials and symbolic regression models across seven observing-system representations. 
    Values above blue bars give RMSD and selected expression complexity (count of operations and constants). 
    All systems use the same 85/15 split.}
    \label{fig:ablation}
\end{figure}

\Cref{fig:ablation} presents the RMSDs of the different models (regressed and discovered) for different cases as indicated by the horizontal axis. 
For the OC3--OC5 models, the linear regression and discovered models perform similarly to each other with almost identical RMSD values between 0.291 and 0.294 on a log$_{10}$ scale.
The OC6 model improves upon the maximum band ocean color models, where both the fitted polynomial and symbolic regression model achieved an RMSD of 0.263 and 0.264, respectively.

For the hyperspectral dataset, the RMSD does not decrease monotonically with an increasing number of input reflectances.
In particular, the 30 bands model with 10 nm resolution achieves the lowest error of approximately 0.232, followed by the 59-band, 5~nm representation with an RMSD of 0.239.
The complete 119-band, 2.5~nm PACE-like representation achieved an RMSD of 0.242.
This suggests that a hyperspectral sensor offers great benefit to building enhanced ocean color models of $\chla$, however, the wavelengths required to do so are sparse and specific.

To further highlight these differences, \cref{tab:oc4_mbr_symbolic_rMSD_relative} outlines the RMSD of each method alongside the percentage change relative to the OC4 baseline, which is the mostly commonly adopted operational model. 
The OC4 baseline achieves an RMSD of 0.291, and is used as the reference in the following comparison.
The symbolic OC3--6 models achieve a relative RMSD of approximately (-0.05\%), (-0.19\%), (-0.01\%), and (-9.15\%), respectively. 
The 10, 5, and 2.5~nm symbolic models reduces the RMSD by approximately (20.41\%), (18.04\%), and (16.91\%), respectively. 
These results further illustrate the benefit of hyperspectral observations, where novel ocean color models can access more informative spectral regions to improve upon standard multispectral ocean color models.

\begin{table}[!htbp]
\small
  \centering
  \caption{Comparison of the OC4 maximum-band-ratio polynomial baseline and symbolic regression RMSD across observation systems. Relative changes are computed with respect to the OC4 maximum-band-ratio polynomial RMSD baseline.}
  \label{tab:oc4_mbr_symbolic_rMSD_relative}
  \begin{tabular}{lcc}
    \hline
    \shortstack{Model\\comparison} & RMSD & \shortstack{Percentage\\change} \\
    \hline
    OC4 MBR polynomial & 0.291 & $0.000\%$ \\
    OC3 symbolic & 0.291 & $-0.05\%$ \\
    OC4 symbolic & 0.291 & $-0.19\%$ \\
    OC5 symbolic & 0.291 & $-0.01\%$ \\
    OC6 symbolic & 0.264 & $-9.15\%$ \\
    10 nm symbolic & 0.232 & $-20.41\%$ \\
    5 nm symbolic & 0.239 & $-18.04\%$ \\
    2.5 nm symbolic & 0.242 & $-16.91\%$ \\
    \hline
  \end{tabular}
\end{table}
  
The ablation results show that retrieval skill does not improve monotonically with the number of available reflectance bands and that a coarser but appropriately placed spectral representation can outperform the full PACE-like input system. 
We next examine whether a non-optical environmental variable, SST, supplies additional information beyond that contained in the reflectances.

\subsection{Ancillary SST information}

In recent literature, researchers have hypothesized the importance of SST as an input to retrieval algorithms to better estimate surface $\chla$ \cite{He2024, Wang2025}. 
While SST was shown to be important to the dynamics of $\chla$ over long time periods \cite{Sun2025Frontiers}, it is still unclear how significant SST is for instantaneous surface $\chla$ retrievals from satellite. 
To this end, we use symbolic regression to discover ocean color models for the hyperspectral matchup dataset, focusing on the OC5 matchups in the presence of SST matchups. 
Note that we ran the symbolic regression algorithm for the remaining OCx models and obtained similar conclusions, hence, those results were omitted. 

\begin{figure}[!htbp]
    \centering
    \includegraphics[width=0.99\linewidth]{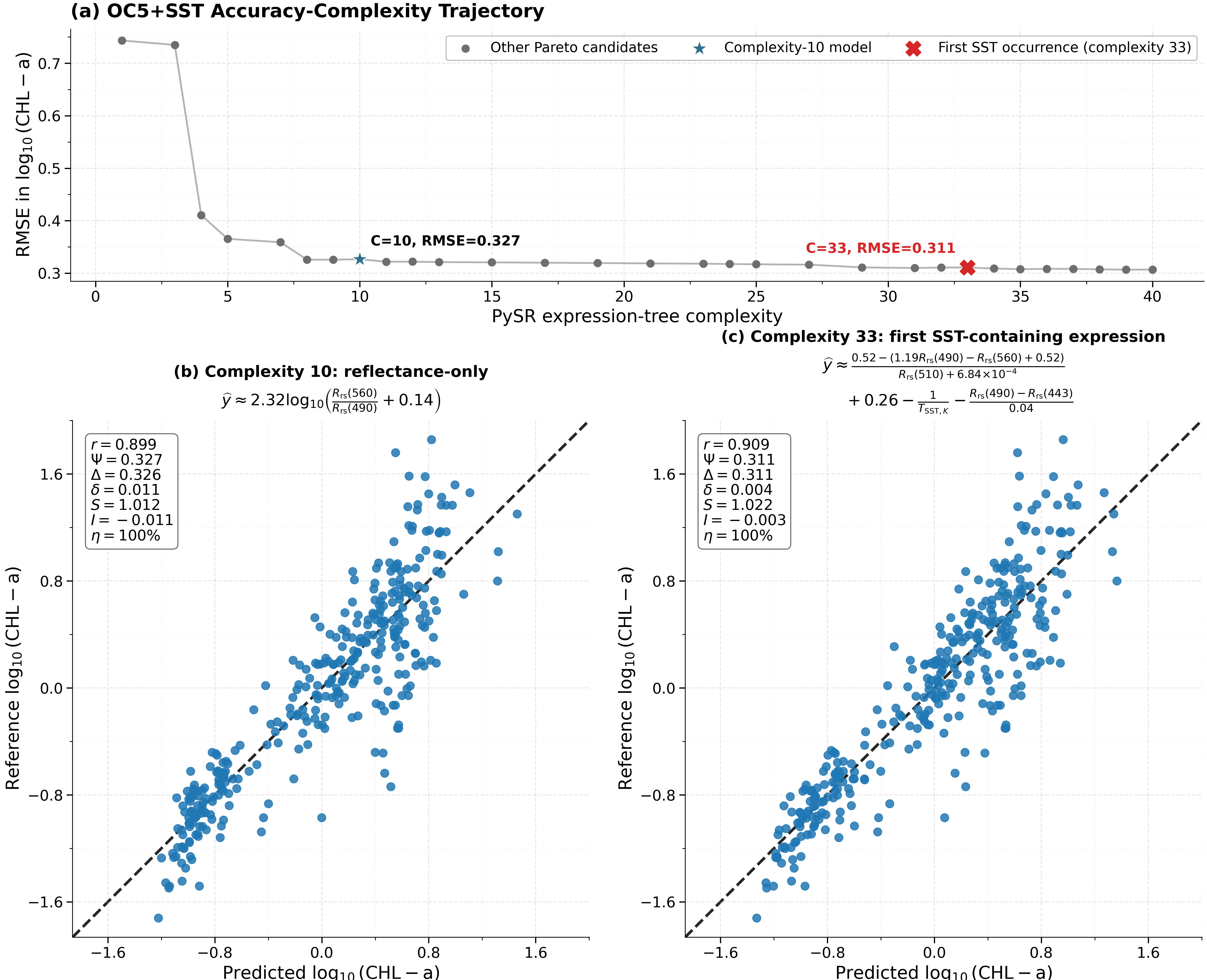}
    \caption{Ancillary SST complexity analysis on one retained OC5 and SST symbolic regression trajectory. 
    (a) Validation selection RMSD across 30 Pareto candidates; the blue star marks the complexity 10 reflectance only model and the red cross marks the first and only SST-containing candidate at complexity 33. 
    (b,c) Parity plots for those two candidates, with display equations rounded and metrics computed from full-precision predictions.}
    \label{fig:sst_complexity}
\end{figure}

\Cref{fig:sst_complexity} summarizes the RMSD--complexity trade-off along the retained discovered models Pareto trajectory and compares the two highlighted candidate expressions. 
In \Cref{fig:sst_complexity}a, the gray circles represent the discovered models, the blue star marks the complexity of 10 and reflectance-only model, and the red cross marks the first and only SST-containing expression, which appears at complexity 33. 
\Cref{fig:sst_complexity}b and \Cref{fig:sst_complexity}c present the corresponding parity plots for the marked complexity 10 and complexity 33 models, respectively. 
The validation RMSD decreases rapidly among the lowest complexity expressions (e.g., 3--4) but plateaus once the complexity reaches approximately 8--10. 
The complexity 10 reflectance-only expression achieves an RMSD of \(0.327\), whereas the complexity 33 SST-containing expression mildly reduces the RMSD to \(0.311\) (approximately \(4.9\%\) improvement), while more than tripling the expression complexity. 
The parity plots show similarly modest improvements in correlation, from \(r=0.899\) to \(r=0.909\), unbiased RMSD, from \(\Delta=0.326\) to \(0.311\), and absolute bias, from \(\delta=0.011\) to \(0.004\). 
Both models retain slopes close to unity, near-zero intercepts, and valid predictions for all observations. 
These results indicate that SST contributes incremental predictive information in this experiment, but the improvement is small relative to the substantial increase in model complexity.

\subsection{Regime diagnostics}

\begin{figure}[htbp]
    \centering
    \includegraphics[width=0.99\linewidth]{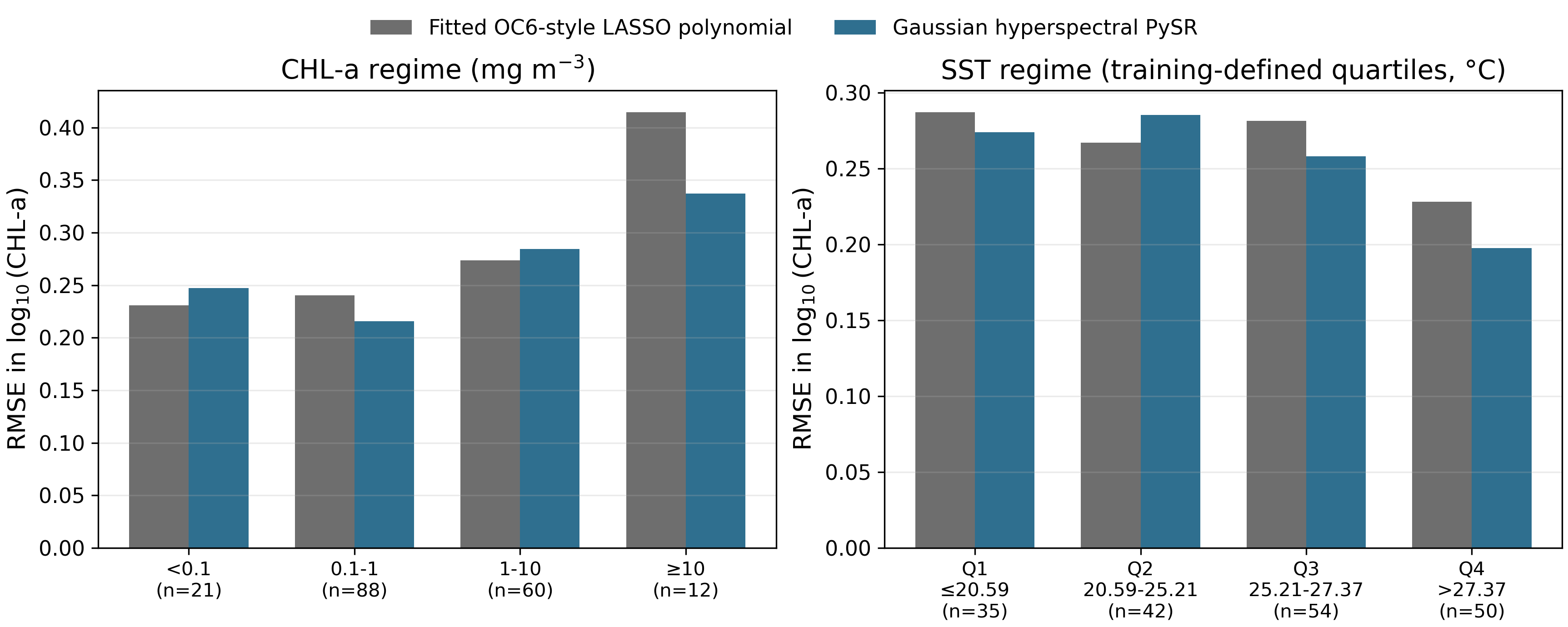}
    \caption{Untouched-test RMSD for the fitted OC6 LASSO polynomial and the exact-complexity 10 PACE-like discovered model across environmental regimes. 
    The left panel stratifies the test observations using fixed $\chla$ concentration ranges, and the right panel uses SST quartiles defined from the training data. 
    Counts beneath the categories indicate the number of test observations in each regime. 
    Neither model was refitted within an individual regime.}
    \label{fig:regimes}
\end{figure}

\Cref{fig:regimes} presents the main regime-dependent error diagnostics for the exact complexity of 10 PACE-like discovered model and the split-fitted OC6 LASSO polynomial. 
The left panel reports the independent test RMSD within four fixed $\chla$ concentration ranges, while the right panel reports the RMSD within SST quartiles defined using the training data. 
Gray bars denote the fitted OC6 polynomial, blue bars denote the PACE-like discovered model, and the labels beneath each category give the number of test samples in the corresponding regime. 
Across the independent test set, the PACE-like model achieved an average RMSD of 0.253, compared with 0.265 for the fitted OC6 model, corresponding to a reduction of 4.65\%.

The strongest variation in performance occurred across the $\chla$ regimes rather than the SST regimes. 
For the PACE-like model, RMSD varied from 0.247 for concentrations below \(0.1~\mathrm{mg\,m^{-3}}\) to 0.216 over \(0.1\)--\(1~\mathrm{mg\,m^{-3}}\), 0.284 over \(1\)--\(10~\mathrm{mg\,m^{-3}}\), and 0.337 for concentrations greater than or equal to \(10~\mathrm{mg\,m^{-3}}\). 
The PACE-like model had lower RMSD than the OC6 model in two of the four concentration regimes, with the largest reduction occurring at \(\ge10~\mathrm{mg\,m^{-3}}\). 
There are two reversals: below \(0.1~\mathrm{mg\,m^{-3}}\) (0.247 versus 0.231) and between \(1\)--\(10~\mathrm{mg\,m^{-3}}\) (0.284 versus 0.274).
Both models exhibited their largest errors in the highest-$\chla$ regime, which often corresponds to turbid conditions that exacerbate the inverse solution, although this bin contained only 12 test observations and should therefore be interpreted cautiously.

Differences across SST regimes were smaller and did not follow a monotonic progression across the quartiles. 
The PACE-like model had lower RMSD than the OC6 model in three of four SST quartiles, with RMSD values of 0.274, 0.285, 0.258, and 0.197 from the coolest to the warmest quartile, respectively. 
The lowest error occurred in the warmest quartile, whereas the second quartile produces the largest RMSD. 
This suggests that the discovered model can outperform standard models in most SST conditions represented in this dataset, rather than only under specific temperature conditions.

The regime analysis characterizes where the models perform well or poorly within held out matchup observations, but it does not establish whether the compact expression can be applied to spatial satellite imagery without introducing discontinuities or unrealistic structures. 
We therefore evaluate its scene-level behavior using two contrasting PACE OCI observations.

\subsection{PACE OCI scene comparison}

\begin{figure}[htbp]
    \centering

    \includegraphics[width=\linewidth]{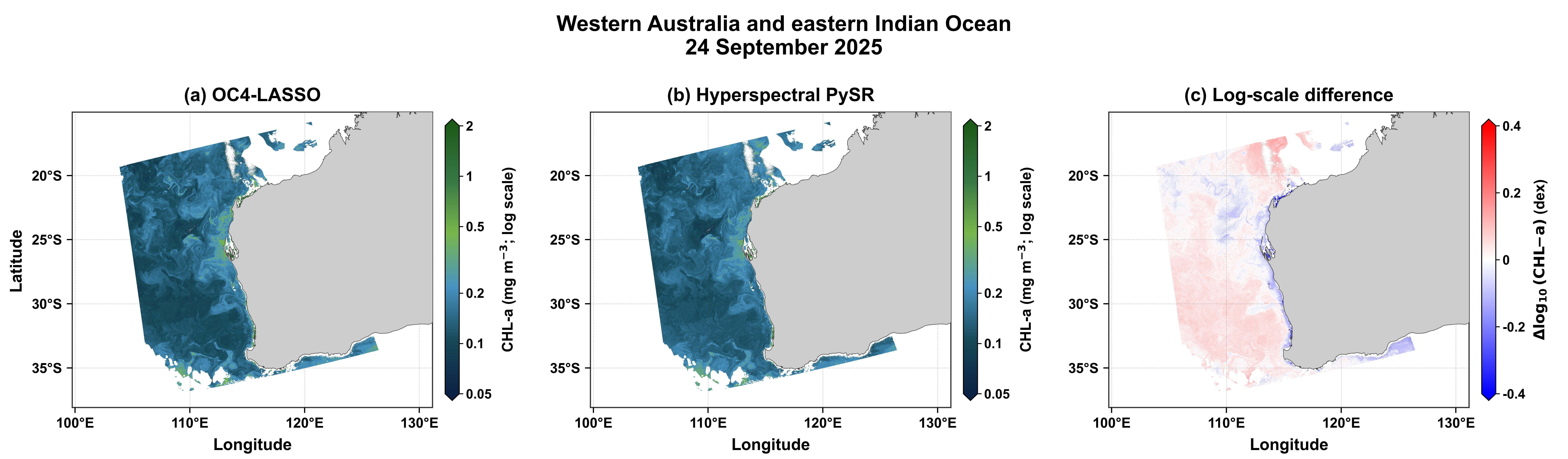}
    \par\medskip
    \includegraphics[width=\linewidth]{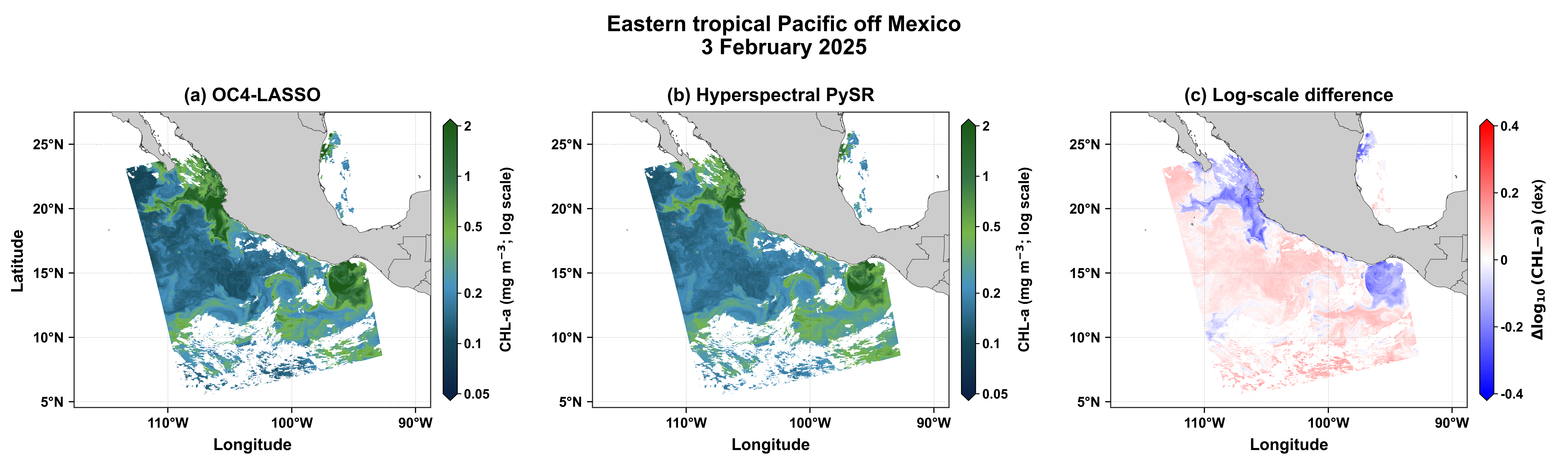}

    \caption{Comparison of fitted OC4-LASSO and hyperspectral PySR chlorophyll-a predictions from PACE OCI observations of Western Australia on 24 September 2025 (upper row) and the eastern tropical Pacific off Mexico on 3 February 2025 (lower row).
    Within each row, the panels show (a) OC4-LASSO, (b) the selected Gaussian-input PySR model, and (c) the logarithmic difference.
    Red indicates higher PySR predictions and blue indicates lower PySR predictions.
    }
    \label{fig:pace_scenes}
\end{figure}

\Cref{fig:pace_scenes} illustrates the application of the discovered expression to two scenes acquired by the hyperspectral Ocean Color Instrument (OCI) aboard PACE.
The first scene examines Western Australia on 24 September 2025, while the second looks at the eastern tropical Pacific off Mexico on 3 February 2025.
For each scene, the left and middle panels show the $\chla$ fields produced by the fitted OC4 LASSO polynomial and the discovered model, respectively.
The right panel shows their pointwise difference in logarithmic space.
Positive differences indicate higher PySR predictions, whereas negative differences indicate higher OC4-LASSO predictions.
Both models were applied to atmospherically corrected OCI remote-sensing reflectances.

For OC4 LASSO, the reflectances were interpolated to the required OC4 wavelengths, whereas for symbolic regression, the hyperspectral inputs were constructed using Gaussian-weighted spectral averaging consistent with the training workflow.
The resulting products exhibit broadly corresponding spatial structures while revealing regional differences in estimated $\chla$.
In the Western Australia scene, both models produce extensive offshore filamentary and eddy-like patterns.
In the Pacific off Mexico, both products exhibit pronounced coastal-to-offshore gradients and coherent offshore structures.
The logarithmic difference fields show spatially organized positive and negative departures, rather than a uniform offset between the products.

The objective of this comparison is to demonstrate the practical feasibility of constructing an alternative satellite-derived $\chla$ product.
The scene-level applications extend the use of the discovered expression beyond the matchup observations used for model development and illustrate its implementation over satellite imagery.
However, no collocated in situ observations were available for these scenes, so a fair comparison would require additional validation data.

A practical advantage of the model discovery approach is that it expresses the relationship between a small subset of hyperspectral inputs and predicted $\log_{10}(\chla)$ in a compact analytical form.
Producing a scene-level field requires evaluating this expression at each valid pixel, supporting straightforward implementation in satellite processing workflows.
The explicit spectral dependence also facilitates inspection of individual band contributions and subsequent sensitivity and uncertainty analyses.
These properties support the development of interpretable alternatives to existing $\chla$ retrievals, and a step towards the next generation of $\chla$ retrievals that exploit state of the art hyperspectral sensors.

\section{Discussion and Conclusions}
\label{sec:discussion}

This study investigated whether symbolic regression can use the expanded spectral information available from PACE-like observations while preserving the compactness and transparency that have made empirical OCx $\chla$ algorithms valuable for ocean color applications. 
The results show that symbolic regression can recover familiar, but not exact, ocean color $\chla$ structure when restricted to conventional multispectral bands and can compress a dense hyperspectral representation into sparse analytical equations. 

The multispectral experiments provide an important bridge to established algorithm design. 
While the symbolic regression algorithm was allowed to explore a broader algebraic search space than the prescribed OCx polynomial family, the selected expressions repeatedly converged toward logarithmic contrasts and ratios involving blue, green, and red reflectances. 
Their predictive skill was generally comparable to that of split-fitted OC3--OC6 polynomials, with modest advantages for some of the larger band systems and modest disadvantages for others. 
More importantly, the discovered equations achieved this performance with fewer fitted constants and an explicit dependence on a small number of individual bands. 
This recurrence of OCx-like structure supports the use of symbolic regression as a transparent and interpretable $\chla$ model discovery method.

When the search was extended to the PACE-like spectral representation, the selected equation used only three wavelengths, combining blue--green and red-edge information in a compact logarithmic ratio. 
The model achieved an unseen test RMSD of 0.253 in log$_{10}$ $\chla$ space with a correlation of 0.939, demonstrating that a sparse equation can retain useful predictive information from a much larger spectral input set. 
The repeated discovery experiments show that across ten outer data splits, the selected symbolic models produced a mean test RMSD of 0.247, compared with 0.307 for the fitted OC6 polynomial. 
The exact equation varied among searches and partitions, but the selected wavelengths repeatedly clustered within the blue--green, green, and red/red-edge regions. 
The robust finding is therefore the recurrence of a sparse spectral family and its predictive advantage, rather than the uniqueness of any one fitted expression.

The observing-system ablation further clarifies how that advantage should be interpreted. Retrieval error did not decrease monotonically as spectral sampling became denser, and the 10~nm representation outperformed both the 5~nm and full 2.5~nm PACE-like systems in the reported experiment. 
Consequently, the benefit of hyperspectral observations should not be equated with using every available band in a retrieval equation. 
Their broader value is that they allow informative spectral neighborhoods to be identified without prescribing them in advance. 
Once identified, those neighborhoods may support simpler selected-band algorithms that are less redundant, easier to calibrate, and more readily transferred across processing systems. 
For the present $\chla$ problem, the results indicate that much of the useful hyperspectral information can be represented by a small number of carefully selected wavelengths. It is important to remember that this analysis is specifically related to empirical $\chla$ ocean colour retrievals. For the retrieval of additional phytoplankon pigments, and information on optically active non-algal components, hyperspectral observations are likely to offer further benefit over multispectral data \cite{Dierssen2023, Kramer2024, Dierssen2021}. 

The SST and regime analyses likewise favor parsimony in the primary retrieval. 
SST entered the symbolic regression trajectory only at substantially higher complexity and reduced RMSD by approximately 4.8\% relative to the compact reflectance-only candidate. 
This modest gain did not offset the more than threefold increase in expression complexity. Differences among SST quartiles were smaller than those among $\chla$ regimes, and the PACE-like model consistently retained lower errors than OC6 across all four SST quartiles in the held-out test. In contrast, errors increased markedly with $\chla$ concentration, with both models performing worst at concentrations above $10~\mathrm{mg,m^{-3}}$. Because this regime contained relatively few observations and is often associated with optically complex waters, additional matchups from high-chlorophyll and turbid waters are needed before strong conclusions can be drawn for these conditions. 

Although SST does not appear to improve algorithm performance when evaluated using discrete matchups, it may become more important when considering broader SST gradients. Notably, the matchup datasets are predominantly located in warmer waters (see Fig. \ref{fig:sst_complexity}), limiting the range of SST represented. Evaluating matchups across contrasting polar and tropical environments, as well as examining longer-term time series in relation to climate variability \citep{Sun2023RSE}, may provide a better assessment of the potential use of SST for satellite $\chla$ retrievals.

The PACE OCI hyperspectral applications demonstrate the practical feasibility of transferring compact symbolic expressions from matchup data to spatial satellite imagery.
In both the Western Australia and eastern tropical Pacific off Mexico scenes, the hyperspectral PySR model produced spatially coherent $\chla$ fields that preserved the principal regional structures while exhibiting systematic local differences from the fitted OC4-LASSO baseline.
These comparisons used common quality and input-validity masks, including training-feature-range screening, and no collocated in situ validation was available.
These pave the pathway towards enhanced $\chla$ estimates at depth \cite{Champenois2024, Champenois2025}.

Overall, symbolic regression offers a useful middle ground between fixed empirical formulas and highly parameterized machine learning models. 
It preserves an analytical form that can be inspected and implemented efficiently, while allowing the data to determine which wavelengths and transformations are most informative. 
Our results suggest that improved $\chla$ retrieval does not necessarily require the full hyperspectral vector, but hyperspectral observations provide the information needed to discover sparse and effective spectral combinations. 
With additional independent validation, for example from PACE, this framework can support the development of interpretable ocean color algorithms for current and future hyperspectral missions. However, achieving this will require continued investment to expand and strengthen in situ datasets.

\section*{Acknowledgements}
A.H. is supported by the Gordon and Betty Moore Foundation’s Postdoctoral Fellowship. 
R.J.W.B and X.S are supported by a UKRI Future Leader Fellowship (MR/V022792/1). 
We would also like to acknowledge high-performance computing support from Princeton University.

\bibliographystyle{unsrtnat}
\bibliography{refs}

\end{document}